\documentclass[conference,twocolumn]{IEEEtran}
\IEEEoverridecommandlockouts
\usepackage{float}
\usepackage{cite}
\usepackage{amsmath,amssymb,amsfonts}
\usepackage{graphicx}
\usepackage{textcomp}
\usepackage{amsmath}
\usepackage{tabularx}
\usepackage{booktabs} 
\usepackage{multirow}
\usepackage[skip=0pt]{caption}
\usepackage{adjustbox}
\usepackage{hyperref}       % hyperlinks
\usepackage{url}            % simple URL typesetting
\usepackage{booktabs}       % professional-quality tables
\usepackage{amsfonts}       % blackboard math symbols
\usepackage{nicefrac}       % compact symbols for 1/2, etc.
\usepackage{microtype}      % microtypography
\usepackage{xcolor}         % colors
\usepackage{algorithm}
\usepackage{soul}
\usepackage{algpseudocode}
\usepackage{amsmath}
\usepackage{tikz}
\usepackage[table]{xcolor}

\usepackage{orcidlink}
\usepackage{comment}
\usepackage{amssymb}
\usepackage{braket}
\usepackage{xcolor}
\newcolumntype{Y}{>{\centering\arraybackslash}X} % centered X column

\def\BibTeX{{\rm B\kern-.05em{\sc i\kern-.025em b}\kern-.08em
    T\kern-.1667em\lower.7ex\hbox{E}\kern-.125emX}}
\begin{document}

\title{FiD-QAE: A Fidelity-Driven Quantum Autoencoder for Credit Card Fraud Detection}

%\author{\IEEEauthorblockN{Anonymous Authors}}

\author{\IEEEauthorblockN{Mansour El Alami\textsuperscript{1}, Adam Innan\textsuperscript{1}, Nouhaila Innan\orcidlink{0000-0002-1014-3457}\textsuperscript{2,3}, Muhammad Shafique\orcidlink{0000-0002-2607-8135}\textsuperscript{2,3}, and Mohamed Bennai\orcidlink{0000-0002-7364-5171}\textsuperscript{1}
}
\IEEEauthorblockA{\textsuperscript{1}Quantum Physics and Spintronic Team, LPMC, Faculty of Sciences Ben M'sick,\\ Hassan II University of Casablanca,
Morocco\\
\textsuperscript{2}eBRAIN Lab, Division of Engineering, New York University Abu Dhabi (NYUAD), Abu Dhabi, UAE\\
\textsuperscript{3}Center for Quantum and Topological Systems (CQTS), NYUAD Research Institute, NYUAD, Abu Dhabi, UAE\\
mansour.elalami-etu@etu.univh2c.ma, adam.innan-etu@etu.univh2c.ma, \\ nouhaila.innan@nyu.edu, muhammad.shafique@nyu.edu,  mohamed.bennai@univh2c.ma \\
}}

\maketitle

\begin{abstract}
Credit card fraud detection is a critical task in financial security, as fraudulent transactions are rare, highly imbalanced, and often resemble legitimate ones. A wide range of classical machine learning methods, as well as more recent quantum machine learning approaches, have been investigated to address this challenge, each providing valuable progress but also leaving open questions regarding scalability, robustness, and adaptability to evolving fraud patterns.  
 In this work, we introduce the Fidelity-based Quantum Autoencoder (FiD-QAE), a quantum architecture that employs fidelity estimation as the decision criterion for anomaly detection. Transactions are encoded into quantum states, compressed through a variational quantum circuit, and evaluated using the SWAP test to distinguish legitimate from fraudulent transactions. We conduct a comprehensive evaluation of FiD-QAE, including statistical analyses, multiple performance metrics, and robustness tests under quantum noise models. The results show that FiD-QAE maintains consistent performance across different imbalance levels and preserves robustness in noisy conditions. Moreover, validation on IBM Quantum hardware backends confirms the feasibility of our approach on real devices, with outcomes consistent with simulation.  
These findings position quantum fidelity as a powerful criterion for anomaly detection and highlight FiD-QAE as a promising direction that complements existing classical and quantum approaches, offering robustness and generalizability for financial fraud detection in realistic environments.  
\end{abstract}


\section{Introduction}
In the modern world, the rapid development of digital technologies, combined with the massive growth of online transactions, has profoundly transformed global payment systems. Among these, credit card payments occupy a central place, both for consumers and financial institutions. This development has been accompanied by an alarming increase in fraudulent activity, posing a significant challenge to the modern financial system.
The consequences are severe for both financial institutions and consumers, resulting in significant economic losses and undermining public confidence in payment systems \cite{awoyemi2017credit}.
According to Nilson Report \cite{nilson2025_cardfraud2023}. In 2023, losses due to credit card fraud reached \$33.83~billion worldwide, compared to \$33.43~billion in 2022, while a joint assessment by the European Banking Authority and the European Central Bank indicated that credit card fraud reached 633 million euros in the first half of 2023 \cite{fraud2024}.
Meanwhile, in the United States, the FBI reported that total losses due to online fraud in 2024 amounted to £16.6 billion, an increase of 33\% compared to 2023\cite{FBI2025}.

Although the financial sector has witnessed significant growth in innovation, particularly through the adoption of artificial intelligence and machine learning (AI/ML) techniques \cite{ryll2020transforming}, traditional approaches remain limited. While often effective, they struggle to handle the complexity and scale of financial data, provide near real-time detection, and adapt to the continuous evolution of fraud strategies. Fraudulent schemes are becoming increasingly sophisticated and dynamic, frequently outpacing these established defense systems \cite{obeng2024utilizing, kulatilleke2022challenges, mienye2024deep, alarfaj2022credit, wawge2025survey}. This underscores the urgent need for more robust, adaptive, and intelligent solutions capable of identifying fraudulent behavior in a rapidly digitizing world.
\begin{figure*}[htpb]
    \centering
    \includegraphics[width=1\linewidth]{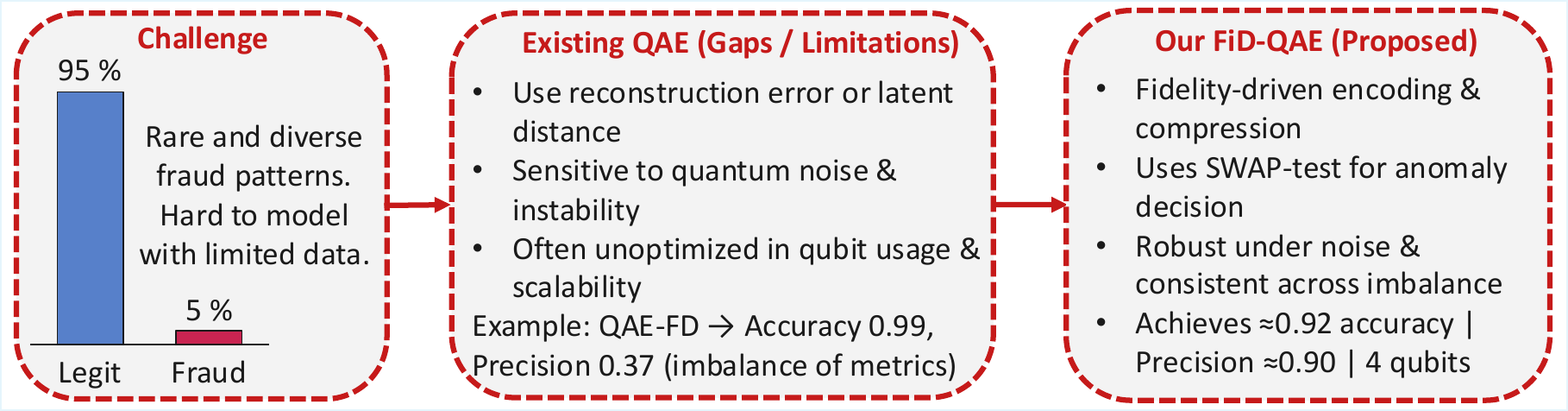}
    \caption{Motivational flow illustrating the reasoning from data imbalance challenges to our proposed FiD-QAE architecture. The process begins with the difficulty of detecting rare and diverse fraud patterns under highly imbalanced datasets, moves through the limitations of existing quantum autoencoders that rely on reconstruction-based detection and exhibit instability, noise sensitivity, and metric imbalance, and culminates in our proposed approach (FiD-QAE), which employs fidelity-driven encoding and SWAP-test evaluation to achieve stable, quantum-consistent anomaly detection and robustness under noise using an efficient 4-qubit design.}
    \label{fig:placeholder}
\end{figure*}
In this context, quantum machine learning (QML),  an emerging paradigm that integrates classical ML with quantum computing (QC) \cite{biamonte2017quantum,schuld2021machine,pineda2025integrating}, offers promising opportunities \cite{innan2023enhancing,innan2024variational,el2024quantum,innan2025optimizing,innan2025qnn,dave2025sentiqnf,innan2024quantum,dutta2025quiet}. Rather than replacing classical methods, QML is envisioned as a complementary approach \cite{ishtiaq2024quantum}, leveraging quantum phenomena such as superposition and entanglement to address existing limitations. These phenomena enable QML to capture complex correlations in large-scale financial datasets and facilitate near real-time classification \cite{liu2018quantum, pistoia2021quantum,innan2024lep,dutta2024qadqn,pathak2024resource,choudhary2025hqnn}. However, the field is still in its early stages, and much remains to be understood about how to translate and exploit these quantum effects effectively \cite{schuld2022quantum,bowles2024better,dutta2025qas}. Despite this, QML holds strong potential to complement conventional ML models and enhance fraud detection accuracy, making it a promising direction to explore given the substantial progress already achieved in the field.

Building on this progress, several supervised quantum models, such as variational quantum circuits (VQCs), and quantum neural networks (QNNs) have been proposed as promising alternatives to classical methods \cite{innan2025next}. Although they show theoretical potential, their effectiveness in real-world situations is limited by structural constraints. These models rely on the availability of balanced labeled data, a condition rarely met in fraud datasets. Furthermore, they are particularly sensitive to barren plateaus, quantum noise, and optimization instability, which makes them difficult to scale and limits their effectiveness in highly variable real-world environments.

To address some of these challenges, the Quantum Autoencoder (QAE) was introduced by Romero et al. \cite{romero2017quantum} as a promising approach for anomaly detection, including financial fraud detection. As part of the unsupervised learning paradigm, the QAE leverages an architecture capable of efficiently compressing quantum data into a latent space while preserving essential information, and subsequently reconstructing quantum states. By exploiting the properties of quantum circuits, QAEs can enhance the identification of anomalies, which are defined as patterns or observations that deviate from expected system behavior \cite{zamanzadeh2024deep}. Such deviations may indicate critical events such as malfunctions, policy violations, or system failures, making anomaly detection a key requirement in domains like credit card fraud prevention. Empirical research has demonstrated the effectiveness of QAEs in detecting anomalies across multiple application areas, including financial fraud \cite{huot2024quantum}, medical anomaly detection \cite{frehner2025applying}, and network security \cite{hdaib2024quantum}, with encouraging results reported in recent studies \cite{bordoni2023long}.

Despite these advances, detecting financial fraud remains particularly challenging due to the extreme imbalance and variability of fraudulent transactions. As shown in Fig.~\ref{fig:placeholder}, existing approaches, both classical and quantum, struggle to address these issues effectively, often exhibiting instability, reconstruction bias, and noise sensitivity. This imbalance motivates the shift toward a fidelity-driven perspective, where anomalies are identified based on the quantum state similarity rather than reconstruction accuracy. Such fidelity-based reasoning offers a more stable, noise-tolerant foundation for quantum anomaly detection in complex financial systems.

In this work, we propose a FiD-QAE architecture for financial fraud detection. The model employs amplitude embedding to encode each transaction into a quantum state and learns to compress normal data into a latent space. Compression fidelity is evaluated using the SWAP test, which compares the discarded (trash) state to a reference state. Since fraudulent transactions lie outside the distribution of training data, they yield poor compression quality, making them distinguishable as anomalies.

This framework provides a flexible solution for detecting rare anomalies in complex financial systems. It is inherently robust to imbalanced datasets, as it focuses on modeling normal transactions. Fraud detection is performed through quantum fidelity, measured via the SWAP test, offering a direct and reliable criterion. Moreover, compression into a reduced subspace mitigates noise, simplifies circuit design, and improves generalization, even under corrupted or previously unseen data.

\textbf{The key contributions of this work are outlined below:}

\begin{itemize}
    \item We establish one of the first dedicated studies of QML for financial anomaly detection, with a focus on credit card fraud, and introduce a tailored quantum algorithm to address this critical challenge.  
    \item We propose a novel Fidelity-based Quantum Autoencoder (FiD-QAE) architecture that exploits only the encoder and compression stages, providing an efficient and scalable quantum framework for fraud detection.  
    \item We present an extensive statistical evaluation on real-world financial datasets, showing that FiD-QAE delivers competitive accuracy while maintaining robustness against data imbalance.  
    \item We demonstrate the practical feasibility of FiD-QAE through preliminary quantum hardware experiments, underscoring its potential for deployment on near-term quantum devices.  
\end{itemize}
The rest of the paper is organized as follows: Sec.~\ref{sec2} introduces the necessary background, presenting the architectures of both classical and quantum autoencoders, along with a review of related literature on classical and QML-based approaches to financial fraud detection. Sec.~\ref{sec3} describes our proposed framework, including the architecture of the QAE model, its operating principles, the encoding method, and the parameterized circuit design. Sec.~\ref{sec4} outlines the datasets used, presents the experimental results, and discusses the key findings. Finally, Sec.~\ref{sec5} concludes the paper and highlights potential directions for future research.

\section{Background and Related Work\label{sec2}}
In this section, we outline the fundamentals and general architectures of classical AE and QAE, both of which are employed for data compression and anomaly detection. We then discuss the key challenges in credit card fraud detection and review existing studies that apply classical and quantum approaches to address this problem.
\subsection{Classical AutoEncoder}

\begin{figure}[htpb]
    \centering
    \includegraphics[width=1\linewidth]{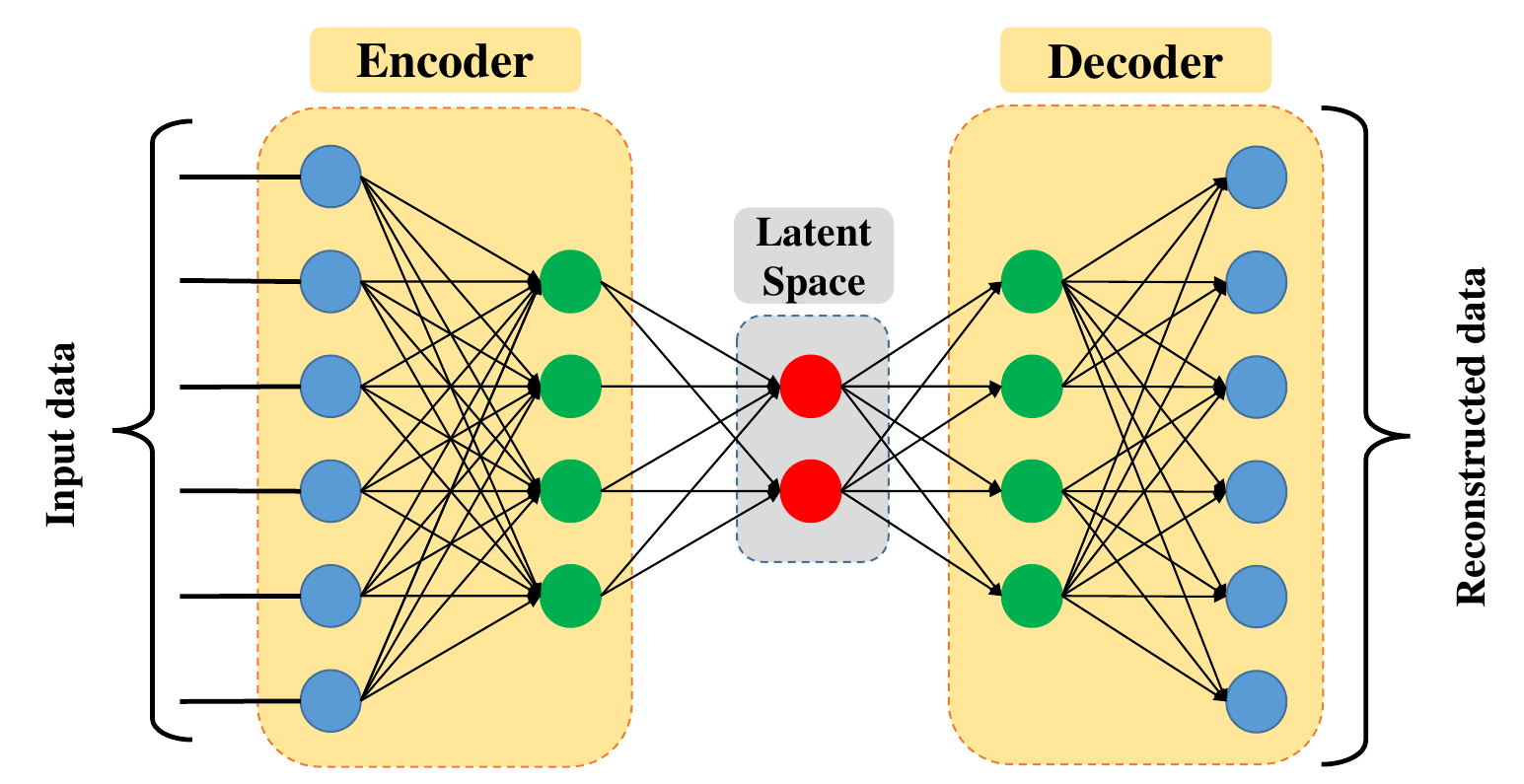}
    \caption{Graphical representation of a classical autoencoder. The encoder compresses the input data into a lower-dimensional latent space, and the decoder reconstructs the input to approximate the original data as closely as possible.}
    \label{fig:CAE}
\end{figure}

Classical AEs are neural networks trained to reconstruct their inputs as accurately as possible \cite{ballard1987modular}. Their primary goal is to learn an embedding representation of the data in an unsupervised manner, which can be applied to various tasks such as anomaly detection and dimensionality reduction \cite{an2015variational,berahmand2024autoencoders}. As illustrated in Fig.~\ref{fig:CAE}, the input data first undergoes an encoding phase, producing a compact latent representation of reduced dimensionality. This is followed by a decoding phase, where the latent representation is used to reconstruct the input data as faithfully as possible.
Classical AEs operate on the principle of jointly optimizing the encoding and decoding processes through iterative training. In this process, data are first passed through the encoder, which generates a latent representation. This representation is then decoded to reconstruct the input. The reconstructed output is compared with the original input, and the reconstruction error is propagated backward through the network to update the encoder and decoder weights using backpropagation. The optimizer continuously adjusts these parameters to minimize the reconstruction error, ensuring that only the most essential structured information is retained \cite{zhou2017anomaly, ng2011sparse, li2023comprehensive}.

\subsection{Quantum AutoEncoder}

\begin{figure}[htpb]
    \centering    
    \includegraphics[width=1\linewidth]{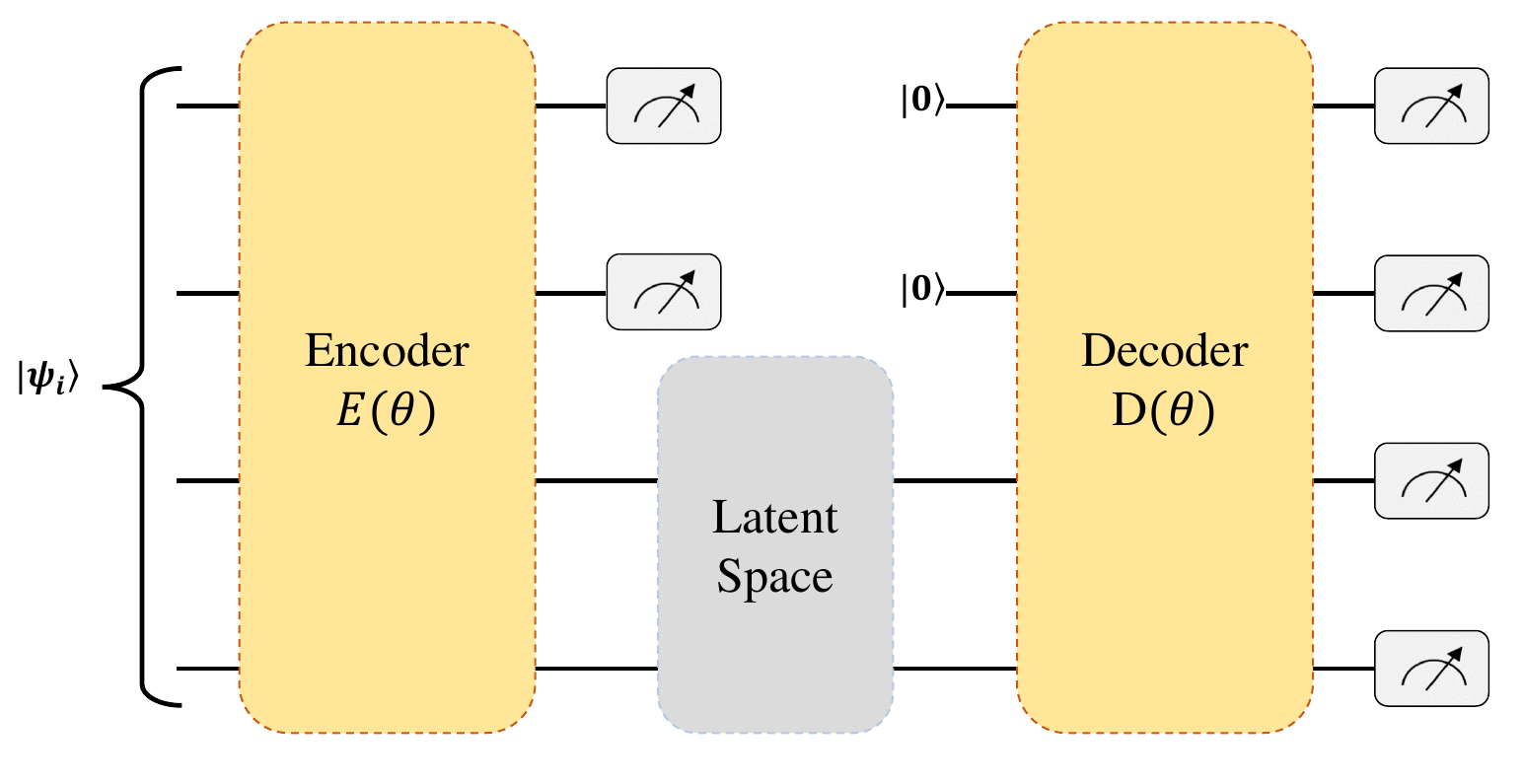}
\caption{Block diagram of the QAE. The model processes four input states, encodes them into two compressed latent states and two trash states, and reconstructs the original four states at the output.}
    \label{fig:enter-label}
\end{figure}

The QAE can be regarded as the quantum analogue of the classical AE. Similar to its classical counterpart, the QAE aims to reduce the dimensionality of the input, which in this case is a quantum state. As illustrated in Fig.~\ref{fig:enter-label}, the QAE architecture consists of two main components: an encoder $E$ and a decoder $D$. The encoder encodes the input quantum state within a parameterized circuit (Ansatz), projecting it into a lower-dimensional latent space. The decoder then uses this compressed state to reconstruct the original state. 

To formalize the quantum encoder, we define two quantum subsystems, $A$ and $B$, containing $n$ and $k$ qubits, respectively. We also introduce a reference space $B'$, associated with a fixed reference state $\ket{a}_{B'}$, often chosen as the ground state $\ket{0}^{\otimes k}$. Let $\ket{\psi}_{AB}$ denote the state of the composite system $AB$, containing a total of $n+k$ qubits.  

The objective is to transform $\ket{\psi}_{AB}$ into a state of the form $\ket{\phi}_A \otimes \ket{\text{trash}}_B$, where the useful information is preserved in subsystem $A$, while $B$ is disentangled and mapped to an input-independent reference state. This is achieved through an encoding operation $E(\theta)$, parameterized by a set of trainable variational parameters $\theta$:  
\begin{equation}
    E(\theta)\left( \ket{\psi}_{AB} \right) = \ket{\phi}_{A} \otimes \ket{\text{trash}}_{B}.
    \label{eq:encoded}
\end{equation}

This operation must disentangle the two subsystems so that $B$ loses all correlation with $A$ and can be discarded. To reconstruct the original state, a quantum decoding operation $D(\theta)$ is applied, where $D(\theta) = E(\theta)^\dagger$, ideally reversing the encoding process. Applying the decoder to the compressed state then reconstructs the original state:
\begin{equation}
D(\theta)\left(\ket{\phi}{A} \otimes \ket{\text{trash}}{B}\right) = \ket{\psi}_{AB}.
\label{eq:decoded}
\end{equation}

The learning task of the QAE is therefore to identify parameterized unitaries that preserve the quantum information of the input state while using a smaller latent space. This requires measuring the deviation between the input $\ket{\psi_i}$ and the reconstructed output $\rho_i^{\text{out}}$. The performance is quantified by the fidelity \cite{wilde2013quantum}:
\begin{equation}
F\left(\ket{\psi_i}, \rho_i^{\text{out}}\right)=\bra{\psi_i} \rho_i^{\text{out}} \ket{\psi_i},
\end{equation}
where successful autoencoding corresponds to $F \approx 1$.

\begin{figure}[!h]
    \centering
    \includegraphics[width=1\linewidth]{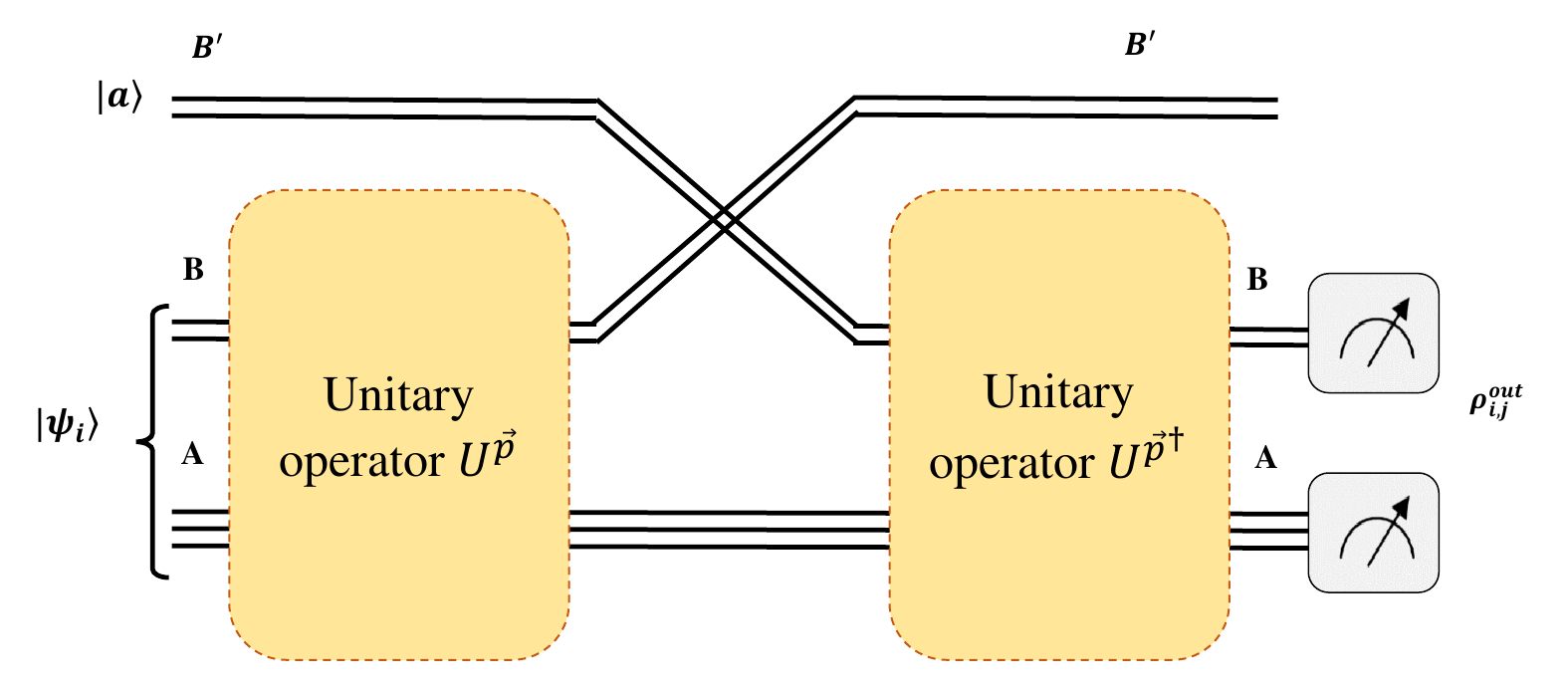}
  \caption{Block diagram of the QAE training process. The objective is to optimize the parameters $\vec{p}$ such that the average fidelity $F\left(\ket{\psi_i}, \rho_i^{\text{out}}\right)$ is maximized.}
    \label{fig:swap_cir}
\end{figure}

\begin{figure}[!h]
    \centering   \includegraphics[width=1\linewidth]{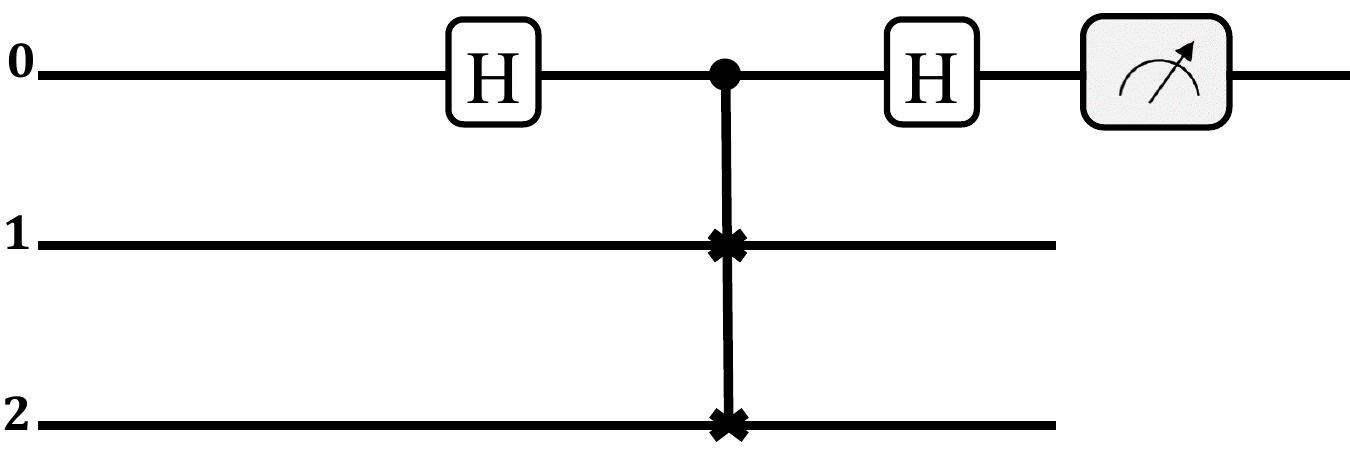}
\caption{SWAP test circuit. The circuit uses a control qubit (qubit 0) initialized with a Hadamard gate, a reference state (qubit 1), a trash state (qubit 2), and a compressed state (qubit 3) to evaluate the fidelity between quantum states.}
    \label{fig:test-swap}
\end{figure}

Formally, let $\{p_i, \ket{\psi_i}_{AB}\}$ denote an ensemble of pure states on $n+k$ qubits, and let $\{U^{\vec{p}}\}$ represent a family of parameterized unitary operators acting on $n+k$ qubits, with $\vec{p} = \{p_1, p_2, \ldots\}$ denoting the variational parameters of the circuit. The cost function to be minimized is the average fidelity:  
\begin{equation}
C_1(\vec{p}) = \sum_i p_i \cdot F\left(\ket{\psi_i}, \rho^{\text{out}}_{i, \vec{p}}\right),
\label{eq:C1}
\end{equation}
where
\begin{equation}
\rho^{\text{out}}_{i,\vec{p}} = \left(U^{\vec{p}}_{AB'}\right)^\dagger
\mathrm{Tr}_B \Big[
U^{\vec{p}}_{AB} \big(\ket{\psi_i}_{AB} \otimes \ket{a}_{B'}\big)
\big(U^{\vec{p}}_{AB}\big)^\dagger
\Big] U^{\vec{p}}_{AB'},
\label{eq:rho}
\end{equation}
with $\ket{\psi_i}\bra{\psi_i}_{AB} = \psi_{i,AB}$ and $\ket{a}\bra{a}_{B'} = a_{B'}$. The goal is to optimize the parameters $\vec{p}$ such that the output state maximizes the average fidelity with the input state. This is illustrated in Fig.~\ref{fig:swap_cir}, where instead of tracing over subsystem $B$, the SWAP test (see Fig.~\ref{fig:test-swap}) is used to compare the compressed and reference states.  

\subsection{Credit card and related work}

Over the past decades, a wide range of classical ML techniques have been applied to credit card fraud detection. Popular models such as support vector machines \cite{kumar2022credit}, random forests \cite{aburbeian2023credit,xuan2018refined,xuan2018random}, logistic regression \cite{alenzi2020fraud}, and naive Bayes classifiers \cite{kiran2018credit,husejinovic2020credit} have demonstrated varying levels of effectiveness \cite{itoo2021comparison}. Beyond these standard algorithms, more advanced approaches have been explored, including gradient-boosted models such as LightGBM \cite{taha2020intelligent}, as well as deep learning architectures like autoencoders \cite{tayebi2025combining} and convolutional neural networks \cite{illanko2022big,karthika2023smart}.  

These advances illustrate the maturity of classical fraud detection research, with numerous studies addressing issues such as class imbalance, feature engineering, and real-time scalability. Nevertheless, challenges remain: fraudulent behaviors are adaptive and dynamic, data volumes continue to grow, and achieving reliable detection with minimal false positives remains difficult.  

Given these limitations, our focus in this work shifts toward QML. While classical approaches provide the foundation and remain widely applied in practice, quantum models explore fundamentally new paradigms that may open promising directions for addressing the evolving complexities of financial fraud detection.

Liang \textit{et al.}~\cite{liang2019quantum} proposed two quantum anomaly detection approaches based on density estimation and multivariate Gaussian distributions, which can be applied to fraud detection. Mitra \textit{et al.}~\cite{mitra2021experiments} introduced a hybrid strategy combining QNNs with classical neural networks. Their study explored two main directions: a quantum–classical neural network model and the use of topological data analysis to reduce noise and improve classification performance. Herr \textit{et al.}~\cite{herr2021anomaly} investigated variational quantum–classical Wasserstein GANs, featuring a hybrid quantum generator and a classical discriminator; when applied to a credit card fraud dataset, the model achieved competitive F1-scores compared to traditional methods. 
Kyriienko \textit{et al.}~\cite{kyriienko2022unsupervised} developed a quantum protocol for anomaly detection in credit card fraud, comparing quantum kernel methods with classical baselines and showing that quantum models can outperform classical ones, particularly as the number of qubits increases.

Building on this, Grossi \textit{et al.}~\cite{grossi2022mixed} applied a quantum support vector machine (QSVM) to real financial data, demonstrating how QML can complement classical methods through novel feature exploration strategies. Wang et al.~\cite{wang2022integrating} proposed a QML framework using an enhanced support vector machine with quantum annealing to detect fraud in unbalanced, time-series online transactions. Their work emphasized the challenges of real-time fraud detection and positioned quantum techniques as promising alternatives for complex business applications. Pena \textit{et al.}~\cite{pena2022fraud} employed data re-uploading techniques to train single-qubit classifiers, achieving performance comparable to classical methods while showing that effective QML can be realized with minimal quantum resources.

Further efforts explored more advanced models. Innan \textit{et al.}~\cite{innan2024financial} evaluated several QML models, including QSVMs and QNNs, for credit card fraud detection, confirming the promise of QML while highlighting scalability challenges. Vuppala \textit{et al.}~\cite{vuppala2024hybrid} introduced a hybrid quantum–classical model based on devastating evolutionary dynamic entities, which, while constrained by hardware limitations, showed reasonable effectiveness on smaller datasets. Innan \textit{et al.}~\cite{innan2024financial1} later proposed a quantum graph neural network for fraud detection, demonstrating improvements compared to its classical GNN counterpart.
Recently, more integrated frameworks have emerged. Huot \textit{et al.}~\cite{huot2024quantum} introduced a fraud detection model based on QAEs, illustrating the adaptability of QML to diverse architectures.

In addition to these representative studies, many other works have explored quantum-based approaches to fraud detection and anomaly detection in the finance sector, each with distinct objectives, architectures, and evaluation strategies \cite{alami2024comparative,innan2025qfnn,sawaika2025privacy,innan2025circuithunt}. This diversity reflects the rapidly growing interest in QML for finance, but also highlights the need for frameworks tailored to specific challenges such as scalability, robustness, and real-time performance. QAEs are particularly attractive in this regard, as they provide efficient compression of quantum states while preserving essential information, with reconstruction fidelity serving as a natural indicator of anomalies. 

However, most existing approaches have not yet been adapted to the unique requirements of financial fraud detection, where data imbalance, evolving patterns, and the need for reliable anomaly identification remain major limitations. Motivated by these developments, we propose the FiD-QAE architecture, which builds on the strengths of QAEs while explicitly addressing these challenges in the context of financial fraud detection.

%%%%%%%%%%%%%%%%%%%%%%%%%%%%%%%%%%%%%%%%%%%%
\section{Methodology\label{sec3}}

\begin{figure*}[htpb]
    \centering
    \includegraphics[width=\linewidth]{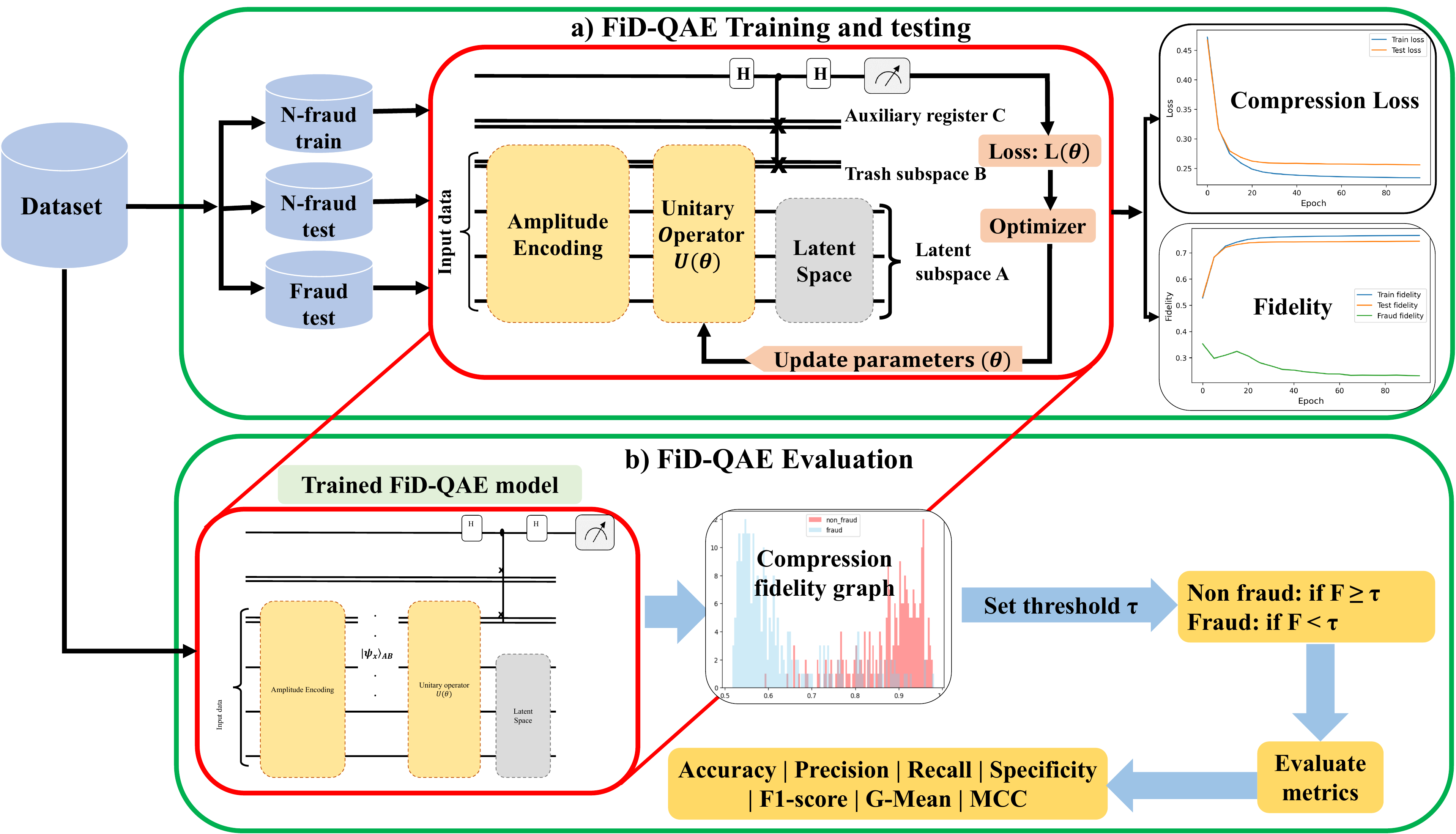}
\caption{Methodology of the FiD-QAE.  \textbf{(a)} Training scheme: an input state $\ket{\psi_i}$ is compressed using a parameterized unitary $U(\theta)$, and fidelity between the reference and trash states is estimated via a SWAP test. The results across all training states define a cost function, which is minimized through classical optimization until convergence, yielding the optimal parameters $\theta=(\theta_1,\theta_2,...)$. \textbf{(b)} Classification workflow: after training, the FiD-QAE is evaluated on new data based on fidelity. A threshold $\tau$ is applied, where transactions with lower fidelity are classified as fraudulent and those with higher fidelity as non-fraudulent. Performance is assessed using standard evaluation metrics.}
    \label{fig:methodology2}
\end{figure*}

\begin{figure}[htpb]
    \centering    \includegraphics[width=1\linewidth]{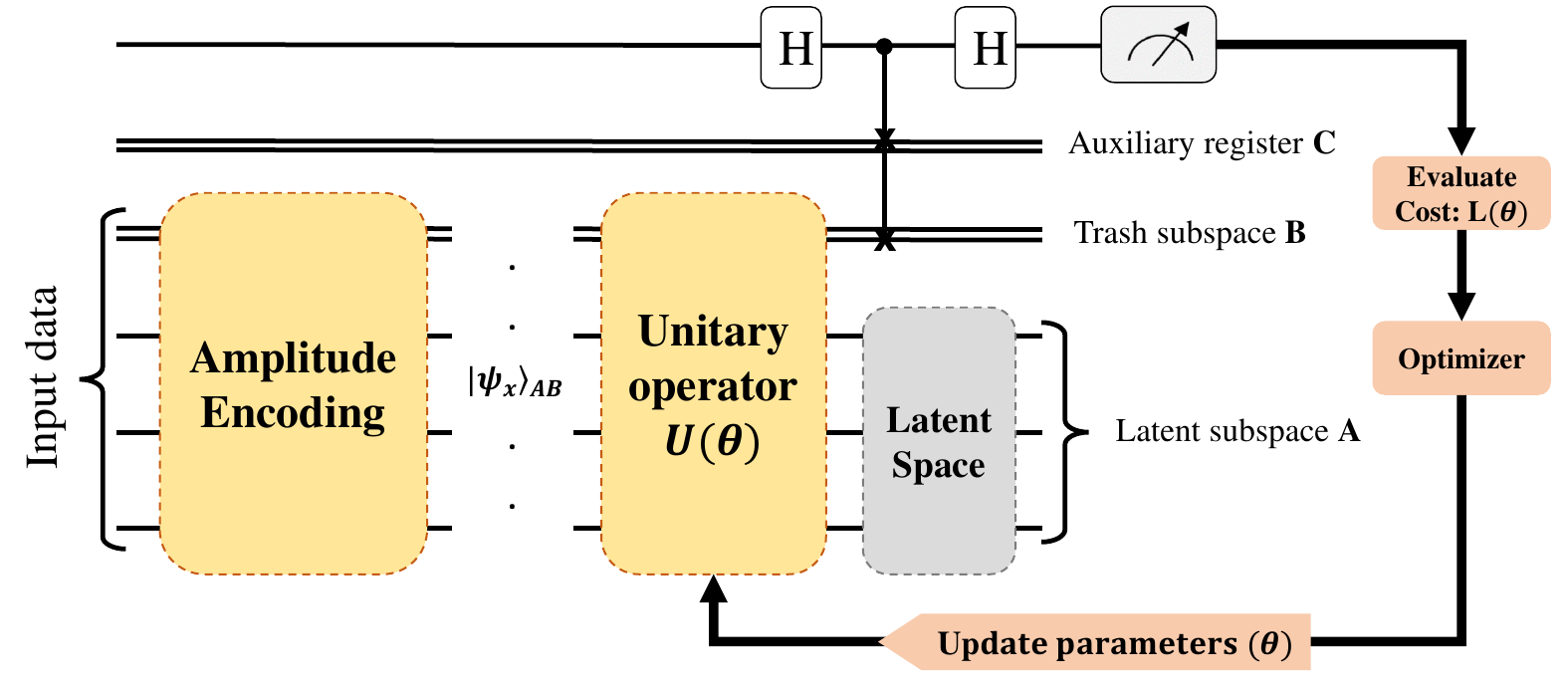}
\caption{Block diagram of the FiD-QAE architecture. Transaction data is encoded into $4$ qubits, compressed into $3$ latent qubits, and one trash qubit is discarded. A SWAP test evaluates the fidelity between the trash qubit and a reference state, which defines the loss function. Circuit parameters $U(\theta)$ are optimized iteratively until convergence.}
    \label{QAE_Model}
\end{figure}

In the FiD-QAE architecture, the input data is preprocessed and normalized before undergoing data encoding; it is then processed by the Quantum Encoder circuit, followed by compression via the SWAP test. The workflow, illustrated in Fig. \ref{fig:methodology2}, integrates the definition of a cost function, optimization, and training, while the overall FiD-QAE structure, composed of two basic blocks, is shown in Fig. \ref{QAE_Model}, with model evaluation performed in the final stage.
\subsection{Data Encoding}
To encode classical data into quantum states for processing, we employ amplitude encoding, which maps a normalized feature vector into the amplitudes of a quantum state. This encoding is well-suited for high-dimensional data and has demonstrated strong representational capacity in other QML tasks \cite{tacchino2019artificial, mangini2020quantum, tacchino2021variational}.

Let a feature vector \(\vec{x}= (x_0, x_1, \ldots, x_{N-1}) \in \mathbb{R}^N\) be normalized such that:
\begin{equation}
    \sum_{i=0}^{N-1} |x_i|^2 = 1.
\end{equation}
It is then encoded as:
\begin{equation}
    |\psi_x\rangle = \sum_{i=0}^{N-1} x_i |i\rangle,
    \label{eq:amplitude_encoding}
\end{equation}
where \(\{|i\rangle\}\) is the computational basis of a register of \(n=\log_2(N)\) qubits.  

The qubits are then divided into two subspaces: latent space \(A\) of size \(n_A\), and trash (ancillary) space \(B\) of size \(n_B\), with \(n=n_A+n_B\). The initial state is therefore:
\begin{equation}
    |\psi_x\rangle_{AB} \in \mathcal{H}_A \otimes \mathcal{H}_B,
\end{equation}
where $\mathcal{H}_A$ and $\mathcal{H}_B$ are the Hilbert spaces of subsystems $A$ and $B$, respectively. An auxiliary register \(C\), initialized as \( |\phi\rangle_C = |0\rangle^{\otimes n_B}\), is later used as a reference state for fidelity measurement.

\subsection{Quantum Encoder Circuit}
The core of the FiD-QAE is the parametric unitary encoder \( U(\boldsymbol{\theta}) \), constructed from $\text{CNOT}$, $R_X$, $R_Y$, and $R_Z$ gates. Its role is to entangle and compress information into the latent space while discarding redundancy into the ancillary space.  

\begin{figure*}[htpb]
     \centering
     \includegraphics[width=\linewidth]{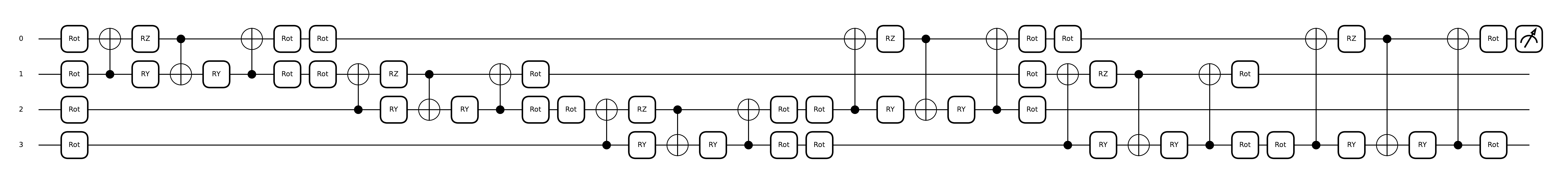}
\caption{Parameterized quantum circuit employed in the FiD-QAE. The design alternates layers of single-qubit rotations and CNOT gates, providing a balance between circuit expressibility and manageable depth.}
     \label{fig:Ansatz}
\end{figure*}

To ensure expressiveness with polynomially bounded depth, we adopt a programmable circuit ansatz, as shown in Fig.~\ref{fig:Ansatz}, consisting of alternating rotation layers and entangling $\text{CNOT}$ gates. This design requires \textbf{$15(n(n-1)/2)$} trainable parameters, which are optimized iteratively to minimize the loss function.

\subsection{Compression via SWAP Test}
After applying $U(\boldsymbol{\theta})$, implicit compression is performed by comparing the sub-space of qubit \textit{trash} with an initial reference state $|\phi\rangle_C = |0\rangle^{\otimes n_B}$, using a SWAP test. This measures the similarity between these two state. The probability of measuring $|0\rangle$ in the control qubit of SWAP test, noted \(P_0\) is related to the quantum fidelity $F$ between reduced state $B$, noted \( \rho_B = \mathrm{Tr}_A \left(|\psi_{\boldsymbol{\theta}}(x)\rangle \langle \psi_{\boldsymbol{\theta}}(x)|\right) \), and reference state \( |\phi\rangle_C \) given as:

\begin{equation}
P_0 = \frac{1}{2} + \frac{1}{2} \mathcal{F}(\rho_B, |\phi\rangle),
\label{proba}
\end{equation}

\begin{equation}
    \mathcal{F}(\theta) = \frac{1}{N} \sum_{i=0}^{N-1} \mathcal{F}\Big(\rho_B(x_i; \theta), \ket{\phi}\Big).
\end{equation}

The fidelity close to \(1\) indicates optimum compression quality, while a low fidelity indicates poor compression quality. 

\subsection{Cost Function}
The cost function, denoted \( L(\theta) \), is defined as the inverse of the fidelity resulting from the SWAP test, as expressed:

\begin{equation}
    \mathcal{L}(\theta) = \frac{1}{N} \sum_{i=0}^{N-1} \Big( 1 - \mathcal{F}\Big(\rho_B(x_i; \theta), \ket{\phi}\Big) \Big).
\end{equation}
This explicitly directs the optimization toward maximizing fidelity; indeed, minimizing this cost function is equivalent to maximizing fidelity. This choice of formulation guarantees a controlled increase in the value of the cost function as fidelity increases, which is perfectly consistent with the overall learning objective.

\subsection{Optimization and Training}
Parameter optimization \(\theta\) is performed using the Adam algorithm, a stochastic gradient descent method with momentum, adapted to the continuous parameters of the quantum circuit. The parameters are updated according to the following equation:
\begin{equation}
    \boldsymbol{\theta}^\ast \leftarrow \boldsymbol{\theta} - \eta \nabla_{\boldsymbol{\theta}} \mathcal{L}(\boldsymbol{\theta}),
\end{equation}
with \(\eta\) as the learning rate.  
The model is trained in the following iterative steps:
\begin{enumerate}
   \item Prepare the input state $\ket{\psi_i}$ and the reference state.
   \item Activate the set of parameters $\theta$ under the unitary encoding $U(\theta)$ at a given optimization step.
   \item Apply a SWAP test to measure the fidelity between the reference state and the trash state.
\end{enumerate}

Once all fidelity values have been estimated, the cost function \(L(\theta)\) is evaluated and passed to the classical optimizer, which outputs an updated set of parameters \(\theta\) for the compression circuit. This process is repeated iteratively until the optimization converges. Algorithm \ref{FiD-QAE-train} goes into detail about how to train and test the FiD-QAE model.

\begin{algorithm}[htpb]
\caption{FiD-QAE}

\label{FiD-QAE-train}

\begin{algorithmic}[1]
\Require Splitting data $D_{Non-fraud}^{Train}$ , $D_{Non-fraud}^{Test}$ and fraud data $D_{Fraud}^{Test}$, number of epochs, Learning rate, FiD-QAE circuit.
\Ensure Trained encoder pentameters, Fidelity history, Loss history, Classification metrics.
\State Initialize quantum device, optimizer, and encoder parameters.
\For{each epoch in training and testing }
    \For{each batch in $D_{Non-fraud}^{Train}$}
        \State Apply Amplitude encoding to put input data into FiD-QAE circuit
        \State Initialize auxiliary qubits
        \State Apply FiD-QAE circuit to specified qubits
        \State Determinate number of trash qubits
        \State Perform SWAP test between trash and auxiliary qubits
        \State compute loss $=1-$ average fidelity. 
        \State Update encoder parameters via Adam optimizer
    \EndFor
    \State Save loss and fidelity values
    \For{each batch in $D_{Test}^{Non-fraud}$ and $D_{Test}^{Fraud}$}
        \State Get input states ready and encoded them as above
        \State Perform SWAP test
        \State Evaluate loss and fidelity values
    \EndFor
\EndFor
\State Plot the loss and fidelity  evaluation curves
\State Save trained parameters
\end{algorithmic}
\end{algorithm}

\subsection{Model Evaluation}
As illustrated in Fig. \ref{fig:methodology2}-b. Once the model has been trained only on transactions considered to be non-fraudulent, this allows it to render a faithful compression of normal behavior. Consequently, when a fraudulent transaction is encoded, the QAE fails to produce a faithful compression, resulting in a significant drop in fidelity. It means that high-fidelity transactions are labeled as non-fraudulent, while low-fidelity transactions are marked as potentially fraudulent transactions. The binary classification rule is defined as:  
\begin{equation}
\text{Label} =
\begin{cases}
\text{Non-fraud} & \text{if } F \geq \tau \\
\text{Fraud} & \text{otherwise }
\end{cases}
\label{eq:label}
\end{equation}
where $\tau$ is the threshold that can be determined empirically. We choose this threshold from the fidelity estimation curves observed on the fraud and non-fraud validation sets. To ensure reproducibility and clarity, the complete FiD-QAE workflow,  including training, fidelity estimation, and classification based on the threshold rule, is summarized in Algorithm~\ref{FiD-QAE_class}.  
\begin{algorithm}[htpb]
\caption{Final evaluation using trained FiD-QAE model}
\label{FiD-QAE_class}
\begin{algorithmic}[1]
\Require Non fraud data $D_{Non-fraud}$, and fraud data $D_{Fraud}$, optimal parameters, and FiD-QAE circuit.
\Ensure Fidelity, Classification metrics.
\For{each sample in $D_{Non-fraud}$ and  $D_{Fraud}$}
    \State Apply the trained FiD-QAE model
    \State Evaluate fidelity
\EndFor
\State Plot fidelity curve distribution
\State Determinate final predictions using the threshold $\tau$
\State Compute classification metrics: Accuracy, Precision, Recall, F1-score, ...
\State plot metrics curves
\end{algorithmic}
\end{algorithm}

\section{Results and Discussion\label{sec4}}
\subsection{Experimental Setup}
We evaluate the FiD-QAE model on a publicly available dataset of credit card transactions from European cardholders \cite{ulb2013}. The dataset contains 284,807 transactions, of which 492 are fraudulent (approximately 0.17\%), making it highly imbalanced. Each transaction includes 30 numerical features: \textit{Time}, \textit{Amount}, 28 anonymized components (\textit{V1}–\textit{V28}) obtained via PCA transformation, and a binary \textit{Class} label indicating fraud (1) or non-fraud (0).  

\begin{figure}[!h]
    \centering
    \includegraphics[width=1\linewidth]{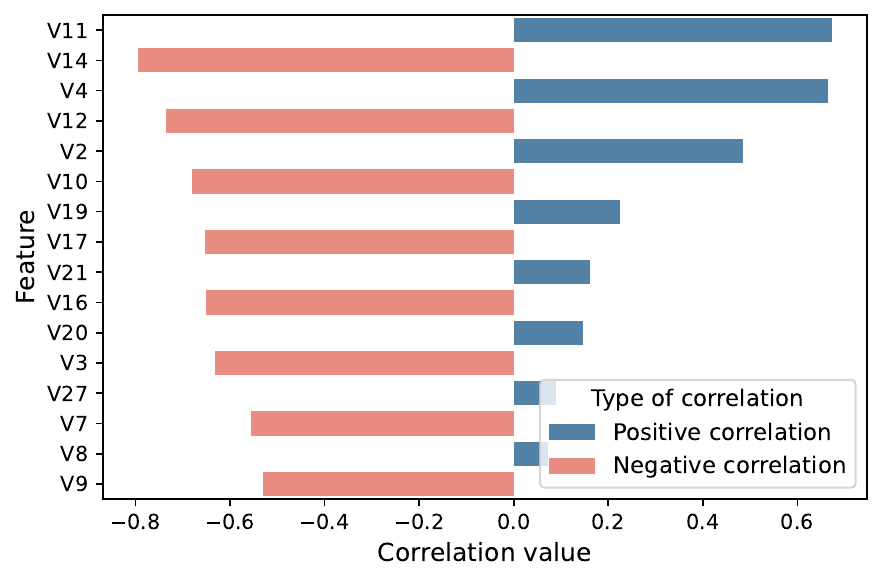}
\caption{Correlation coefficients between the 16 selected features and the fraud label. Notably, features such as ``V11'', ``V14'', and ``V4'' exhibit strong discriminative power.}
    \label{fig:feaut-corr}
\end{figure}

To mitigate the influence of extreme values, the continuous features \textit{Time} and \textit{Amount} are normalized using the \texttt{RobustScaler} from \texttt{scikit-learn}, ensuring consistent scaling while preserving their distributional structure. Since the quantum model can only process a limited number of features through amplitude encoding, a feature selection step is performed. We compute the linear correlation of each feature with the class label and retain the 16 features with the highest absolute correlation values. As illustrated in Fig.~\ref{fig:feaut-corr}, features such as \textit{V11}, \textit{V14}, and \textit{V4} exhibit strong discriminative power, making them particularly relevant for fraud detection. This procedure reduces dimensionality, suppresses quantum noise, and improves the statistical significance of the encoded data.  

The configuration and hyperparameters used in our experiments are summarized in Table~\ref{tab:parameter_values}. The model is trained for 100 epochs with a batch size of 64, using the Adam optimizer with a learning rate of 0.001. To calibrate the model’s sensitivity in distinguishing fraudulent from legitimate transactions, multiple threshold values are explored. The FiD-QAE is implemented using the \texttt{PennyLane} framework \cite{bergholm2018pennylane}, with the \texttt{default.qubit} simulator employed for experimental results and \texttt{Qiskit} backend used for execution on IBM Quantum hardware \cite{javadi2024quantum}.  

\begin{table}[h]
    \centering
    \caption{Model configuration and hyperparameters.}
    \begin{tabularx}{\columnwidth}{YY}
        \hline
        \rowcolor{cyan!20}
        \textbf{Parameter} & \textbf{Value} \\
        \hline
        Number of Qubits        & 4 \\
        Number of Trash Qubits  & 1 \\
        Optimizer               & Adam \\
        Learning Rate           & 0.001 \\
        Batch Size              & 64 \\
        Number of Epochs        & 100 \\
        Threshold Values        & 0.40–0.55, 0.65 \\
        \hline
    \end{tabularx}
    \label{tab:parameter_values}
\end{table}

\subsection{Convergence Analysis}

\begin{figure}[htpb]
    \centering        \includegraphics[width=1\linewidth]{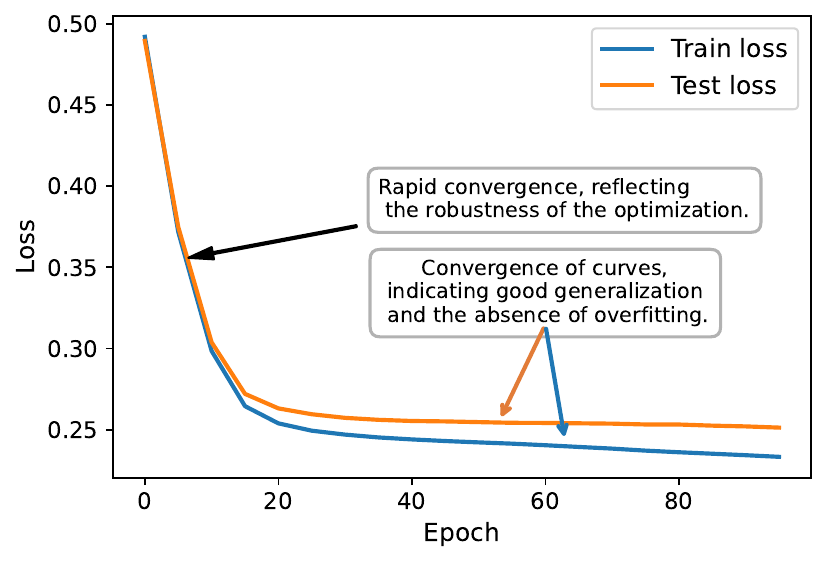}
\caption{Training and testing loss curves of the FiD-QAE on non-fraudulent data.}
    \label{Loss}
\end{figure}

The FiD-QAE model is trained exclusively on non-fraudulent data and evaluated on both non-fraudulent and fraudulent samples, with the objective of maximizing compression fidelity for legitimate transactions while yielding degraded fidelity for fraudulent ones. As shown in Fig.~\ref{Loss}, the model’s loss function, measured only on non-fraudulent data, initially exhibits a rapid and significant decrease, reaching approximately 0.24 within the first twenty iterations. This stage demonstrates effective optimization of the quantum circuit parameters, confirming that the FiD-QAE is capable of quickly extracting compressed representations of non-fraudulent data. As training progresses, the decrease in loss becomes more gradual, stabilizing around 0.23, which indicates convergence to an equilibrium where further updates provide minimal improvement.  
Furthermore, the training and testing curves remain very close throughout the process, highlighting the generalization ability of the FiD-QAE and suggesting that it does not suffer from overfitting, thereby maintaining stable performance on unseen data. This behavior is particularly valuable in fraud detection, where test data may differ in distribution from that observed during training. The consistency of the error on the test set further demonstrates the robustness of the FiD-QAE to fluctuations in input data and indicates that it captures global, discriminative features rather than memorizing specific examples.

\subsection{Fidelity Analysis}

\begin{figure}[htpb]
    \centering
    \includegraphics[width=1\linewidth]{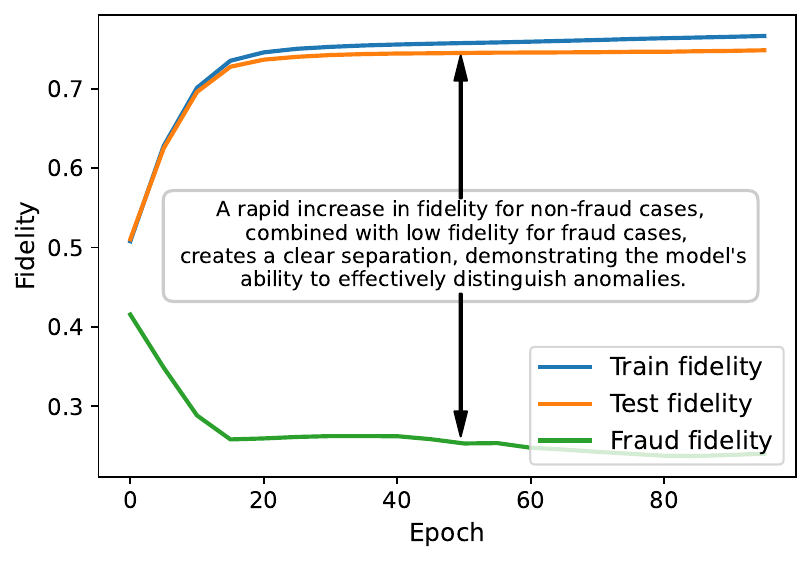}
\caption{Training and testing fidelity curves of the FiD-QAE on fraudulent and non-fraudulent data.}
    \label{Fidelity}
\end{figure}

To provide additional insights into the behavior of the FiD-QAE model, Fig.~\ref{Fidelity} illustrates the evolution of fidelity during training and testing, evaluated on both non-fraudulent and fraudulent data. From the very first epochs, the fidelity on non-fraudulent training data increases rapidly, rising from approximately 0.50 to above 0.76 within the first twenty iterations. This trend indicates efficient optimization of the quantum circuit parameters to achieve high similarity between the trash state and the reference state, which in turn implies good compression quality for legitimate transactions. The fidelity on the non-fraudulent test data follows a similar trajectory, confirming the generalization capability of the architecture on unseen legitimate samples. The evolution of fidelity in this phase closely mirrors the behavior observed in the loss function.  
In contrast, the fidelity evaluated on fraudulent data exhibits the opposite trend. It decreases sharply during the early epochs, falling from around 0.40 to approximately 0.20, and then remains relatively stable for the remainder of training. This behavior is both expected and desirable, as it reflects the FiD-QAE’s ability to recognize deviations from the training distribution. In other words, the model effectively differentiates between legitimate and fraudulent transactions based on fidelity, validating this measure as a reliable indicator for fraud detection.  

These results demonstrate that the FiD-QAE learns compact and relevant representations of transactions, which is essential for effective anomaly detection. The rapid convergence and minimal difference between training and testing curves highlight the robustness of the architecture and its potential for large-scale fraud detection. Furthermore, by maximizing fidelity for non-fraudulent data while driving fraudulent data toward lower fidelity values, the FiD-QAE ensures a clear separation between the two classes.  

%%%%%%%%

\subsection{Threshold-Based Performance Analysis}
\begin{figure*}[htpb]
    \centering    
    \includegraphics[width=\linewidth]{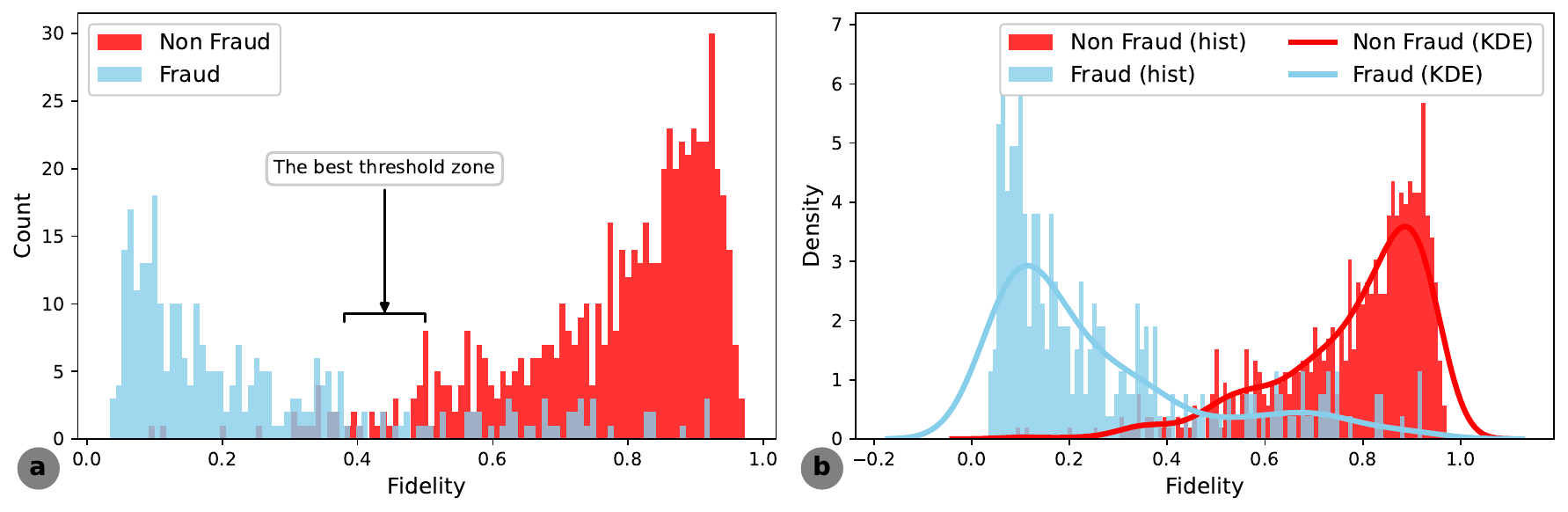}
\caption{Statistical evaluation of quantum fidelity distribution across transaction classes: \textbf{(a)} histogram of fraudulent and non-fraudulent transactions; \textbf{(b)} histogram with KDE.}
    \label{Hist_KDE}
\end{figure*}

The classification ability of the FiD-QAE model is assessed by analyzing fidelity scores after training. As shown in Fig.~\ref{Hist_KDE}, the distribution of fidelity values for non-fraudulent and fraudulent transactions reveals clear separation between the two classes. Subplot Fig.~\ref{Hist_KDE}-a presents raw frequency histograms, while Fig.~\ref{Hist_KDE}-b overlays histograms with kernel density estimation (KDE) to provide a smoother and more interpretable visualization. In subplot \textbf{(a)}, non-fraudulent transactions are concentrated in the high-fidelity range (0.7–1.0), with a pronounced peak around 0.9, confirming the FiD-QAE’s ability to capture the structural characteristics of legitimate data and compress it efficiently.  
The KDE analysis in subplot \textbf{(b)} further emphasizes the separation between classes: non-fraudulent transactions cluster tightly around high fidelity values, while fraudulent transactions accumulate in the low-fidelity zone. Overlap between the two distributions is relatively limited, mainly in the 0.4–0.6 interval, indicating strong discriminative power.  

\begin{figure*}[htpb]
    \centering
    \includegraphics[width=1\linewidth]{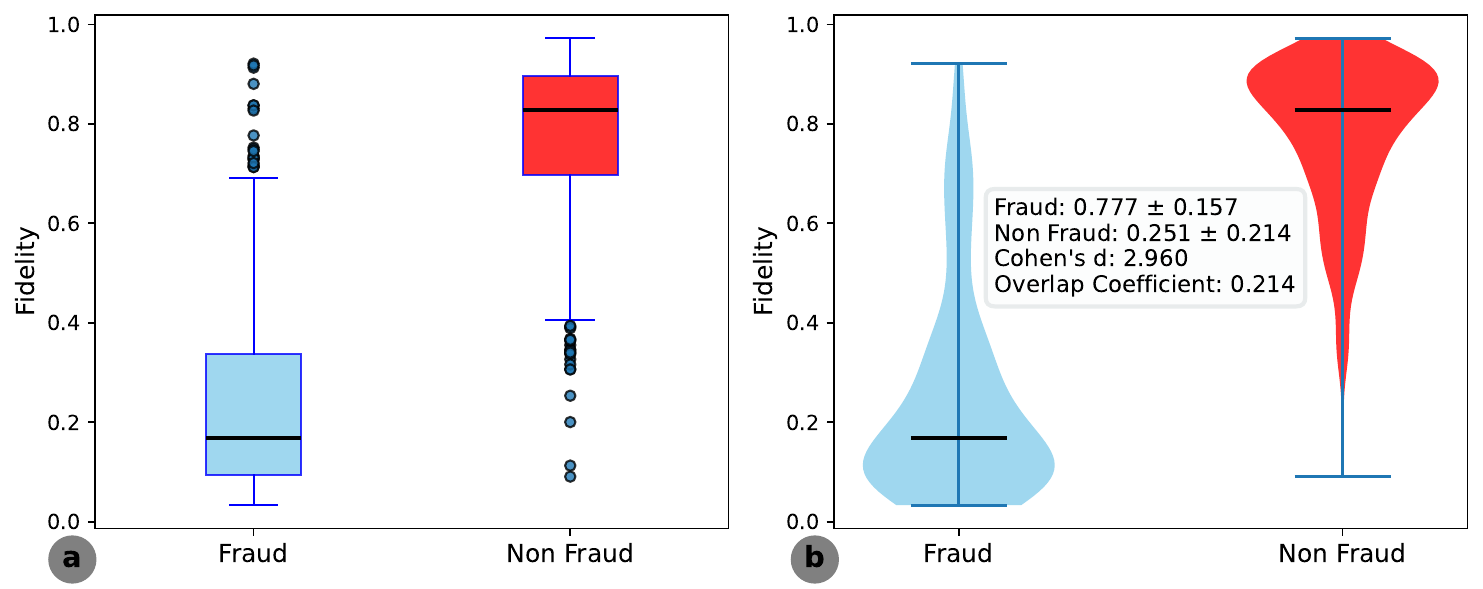}
\caption{Distribution of quantum fidelity for fraudulent and non-fraudulent transactions: \textbf{(a)} box plot illustrating dispersion and extreme values; \textbf{(b)} violin plot representing distribution density with statistical separation indicators (means, Cohen's $d$, and overlap coefficient).}  
    \label{box_violin}
\end{figure*}

To refine this analysis, Fig.~\ref{box_violin} provides statistical comparisons using box plots \textbf{(a)} and violin plots \textbf{(b)}. The box plot shows a clear difference between the two classes: fraudulent transactions average around 0.18, while non-fraudulent transactions average around 0.85. Quartile analysis confirms this separation, with fraud cases clustered between 0.10 and 0.35, and non-fraudulent cases between 0.70 and 0.90. Outliers, however, reveal that a few fraudulent samples achieve relatively high fidelity (suggesting sophisticated attack scenarios), while some legitimate transactions obtain low fidelity (reflecting false positives).  

The violin plot complements this view by illustrating distribution density. Fraudulent transactions show a unimodal density at low fidelity, whereas non-fraudulent transactions exhibit a dominant mode at high fidelity with a downward tail, indicating a few poorly compressed samples. These patterns confirm that although most transactions are clearly distinguished, limited overlap remains.  

\begin{figure}[htpb]
    \centering    \includegraphics[width=1\linewidth]{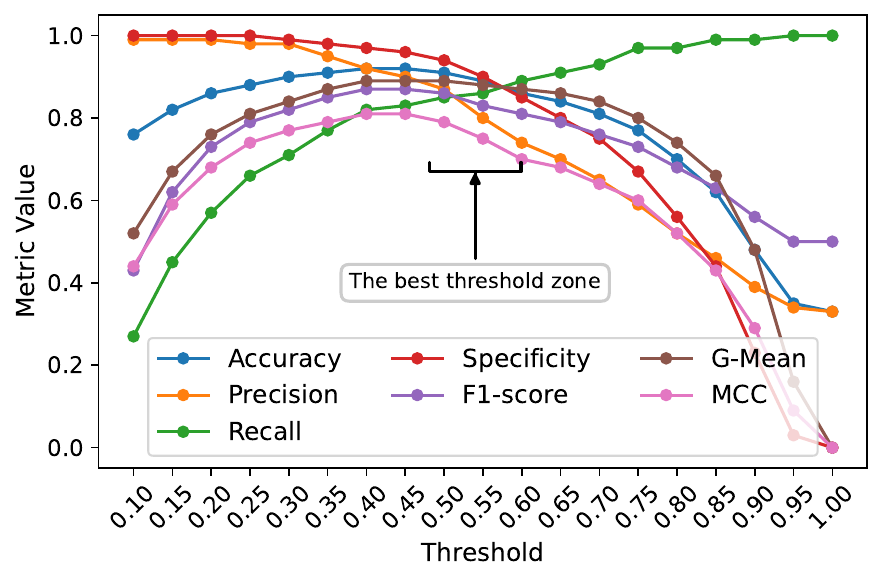}
    %0.8for journal
\caption{Variation of evaluation metrics across different decision thresholds.}
    \label{Metric_Threshold}
\end{figure}

Statistical indicators extracted from Fig.~\ref{box_violin}-a reinforce this distinction. Non-fraudulent transactions achieve an average fidelity of $0.777 \pm 0.157$, while fraudulent transactions average $0.251 \pm 0.214$. The substantial gap between distributions is supported by an exceptionally high Cohen’s $d$ of 9.60, far exceeding conventional thresholds, and an Overlap Coefficient of 0.214, confirming the limited overlap concentrated in the mid-fidelity range.  
To translate these results into operational performance, we define classification thresholds based on the observed fidelity distributions. As shown in Table~\ref{metrics}, the FiD-QAE achieves high accuracy (0.92), near-perfect specificity (0.96–0.97), and stable F1-scores around 0.87 at lower thresholds (0.40–0.45), though recall remains modest (0.82–0.83). This indicates excellent ability to identify legitimate transactions but limited sensitivity to fraud cases. As the threshold increases (0.50–0.55), recall improves (0.85–0.86), but precision and accuracy decrease, reflecting more false positives. Correspondingly, F1-scores drop slightly (0.86 to 0.83), and MCC decreases from 0.79 to 0.75, while the G-Mean remains stable.  
\renewcommand{\arraystretch}{1}
\setlength{\tabcolsep}{2pt}
\begin{table}[htpb]
\caption{Metrics across the optimal threshold interval [0.40, 0.55].}
\centering
\begin{tabular}{>{\centering\arraybackslash}m{1.2cm}
                >{\centering\arraybackslash}m{1.1cm}
                >{\centering\arraybackslash}m{1cm}
                >{\centering\arraybackslash}m{0.7cm}
                >{\centering\arraybackslash}m{1.2cm}
                >{\centering\arraybackslash}m{1.1cm}
                >{\centering\arraybackslash}m{1cm}
                >{\centering\arraybackslash}m{0.6cm}}
\hline
\rowcolor{cyan!20}
\textbf{Threshold} & \textbf{Accuracy} & \textbf{Precision} & \textbf{Recall} & \textbf{Specificity} & \textbf{F1-score}& \textbf{G-Mean}& \textbf{MCC}\\
\hline
0.40 & 0.92 & 0.92 & 0.82 & 0.97 & 0.87 & 0.89 & 0.81 \\
0.45 & 0.92 & 0.90 & 0.83 & 0.96 & 0.87 & 0.89 & 0.81 \\
0.50 & 0.91 & 0.87 & 0.85 & 0.94 & 0.86 & 0.89 & 0.79 \\
0.55 & 0.89 & 0.80 & 0.86 & 0.90 & 0.83 & 0.88 & 0.75 \\
\hline
\end{tabular}
\label{metrics}
\end{table}
As shown in Fig.~\ref{Metric_Threshold}, the variation of performance metrics across thresholds confirms that the FiD-QAE achieves the best trade-off in the intermediate range (0.45–0.50). This balance ensures reliable fraud detection while controlling false positives, an essential requirement for operational deployment in financial systems.  

The FiD-QAE demonstrates high robustness, with particularly strong performance around intermediate thresholds. These findings underscore the importance of threshold selection in achieving an operational balance between maximizing fraud detection and minimizing false positives, a critical requirement for real-world financial fraud detection systems.  

\subsection{Fraud Prevalence Analysis}
Analyzing the robustness of a financial fraud detection model requires assessing its overall performance and examining the impact of the proportion of fraudulent data used in evaluation. To this end, we progressively increased the proportion of fraud cases used in the evaluation of the FiD-QAE model, from 20\% to 80\% of the available fraudulent data, and analyzed the resulting effects on robustness and stability. 

\begin{figure*}[!ht]
    \centering    
    \includegraphics[width=\linewidth]{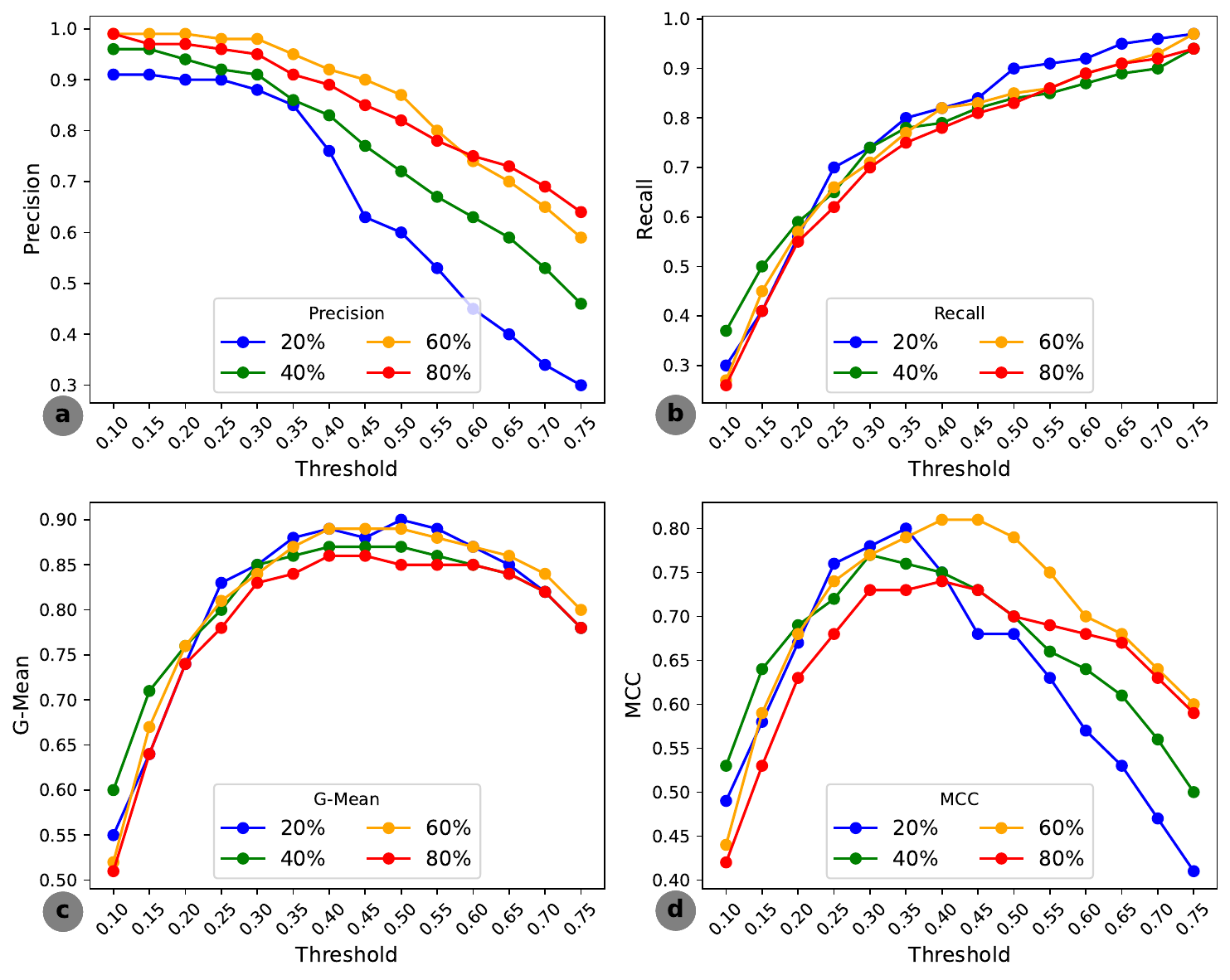}
\caption{Impact of varying fraudulent data proportions (20\%–80\%) on FiD-QAE performance. The curves show the evolution of key classification metrics as a function of the decision threshold: \textbf{(a)} Precision, \textbf{(b)} Recall, \textbf{(c)} F1-score, and \textbf{(d)} MCC, highlighting the overall stability of the FiD-QAE.}
    \label{Fraud_Prevalence}
\end{figure*}

As shown in Fig.~\ref{Fraud_Prevalence}, the main metrics (precision, recall, F1-score, and MCC) are reported as functions of the decision threshold under different fraud prevalence settings. Fig.\ref{Fraud_Prevalence}-\textbf{(a)} presents accuracy curves, where a prevalence of 80\% leads to slightly lower performance at higher thresholds, though accuracy remains satisfactory overall. Precision, however, improves significantly with higher fraud prevalence (40\%–80\%), becoming more consistent and stable across thresholds. This indicates that the FiD-QAE effectively leverages additional fraudulent cases to enhance reliability. In contrast, recall, shown in Fig.\ref{Fraud_Prevalence}-\textbf{(b)}, increases steadily with the threshold and exhibits similar behavior across prevalence levels. The relative proximity of the curves indicates that the model consistently identifies fraudulent transactions regardless of the proportion of fraud in the dataset.  

The F1-score, illustrated in Fig. \ref{Fraud_Prevalence}-c, confirms this balance between precision and recall. While the 20\% scenario yields slightly lower values, performance improves notably at 40\% and peaks around 0.87 at 60\%. Even at 80\%, the FiD-QAE maintains high F1-scores, confirming its effectiveness across different prevalence rates. Finally, the MCC curves in Fig. \ref{Fraud_Prevalence}-d) show the same stability: although performance is slightly lower at 20\%, MCC remains above 0.80 in all cases, with high and consistent values at 40\%, 60\%, and 80\%. The convergence of curves demonstrates the robustness and generalizability of the FiD-QAE model, even under scenarios with varying levels of imbalance.  

Table~\ref{metrics_optimal_threshold} complements this analysis by reporting the optimal values of precision, recall, F1-score, and MCC for each fraud prevalence, considering the best corresponding threshold. We observe that the optimal threshold remains stable between 0.30 and 0.40 across all prevalence levels, which is an important asset for practical deployment. Furthermore, the FiD-QAE consistently achieves high performance, with all metrics above 0.80, regardless of prevalence. These results confirm that variations in the proportion of fraudulent cases do not undermine the model’s reliability. Interestingly, the 60\% prevalence scenario appears most favorable, producing slightly higher F1-scores.  

These results highlight the stability and effectiveness of the FiD-QAE model in the presence of varying fraud rates. The existence of a nearly constant decision threshold, together with the model’s ability to maintain high performance across a wide range of prevalence scenarios, demonstrates its adaptability to real-world financial environments, where data imbalance conditions frequently change. This robustness to heterogeneous distributions represents a major advantage for practical deployment.

%%%%%%%%%%%%%%%

\renewcommand{\arraystretch}{1}
\setlength{\tabcolsep}{2pt}
\begin{table}[htpb]
\caption{Variation of evaluation metrics across different decision thresholds.}
\label{metrics_optimal_threshold}
\centering
\setlength{\tabcolsep}{4pt}
\begin{tabular}{>{\centering\arraybackslash}m{1.5cm}
                >{\centering\arraybackslash}m{1.6cm}
                >{\centering\arraybackslash}m{1cm}
                >{\centering\arraybackslash}m{0.7cm}
                >{\centering\arraybackslash}m{1.2cm}
                >{\centering\arraybackslash}m{1.1cm}
                >{\centering\arraybackslash}m{1cm}
                >{\centering\arraybackslash}m{0.6cm}}
\hline
\rowcolor{cyan!20}
\textbf{Splitting fraud data} & \textbf{Threshold $\tau$} & \textbf{Precision} & \textbf{Recall} & \textbf{F1-score} & \textbf{MCC} \\
\hline

% Bloc 20%
\multirow{3}{*}{\cellcolor{orange!20}\textbf{20\%}}
 & \cellcolor{orange!20}0.30 & \cellcolor{orange!20}0.88 & \cellcolor{orange!20}0.74 & \cellcolor{orange!20}0.80 & \cellcolor{orange!20}0.78 \\
 & \cellcolor{orange!20}0.35 & \cellcolor{orange!20}0.85 & \cellcolor{orange!20}0.80 & \cellcolor{orange!20}0.82 & \cellcolor{orange!20}0.80 \\
 & \cellcolor{orange!20}0.40 & \cellcolor{orange!20}0.76 & \cellcolor{orange!20}0.82 & \cellcolor{orange!20}0.79 & \cellcolor{orange!20}0.75 \\[2pt]

% Bloc 40%
\multirow{3}{*}{\cellcolor{blue!15}\textbf{40\%}}
 & \cellcolor{blue!15}0.30 & \cellcolor{blue!15}0.91 & \cellcolor{blue!15}0.74 & \cellcolor{blue!15}0.82 & \cellcolor{blue!15}0.77 \\
 & \cellcolor{blue!15}0.35 & \cellcolor{blue!15}0.86 & \cellcolor{blue!15}0.78 & \cellcolor{blue!15}0.82 &\cellcolor{blue!15}0.76 \\
 & \cellcolor{blue!15}0.40 & \cellcolor{blue!15}0.83 & \cellcolor{blue!15}0.79 & \cellcolor{blue!15}0.81 & \cellcolor{blue!15}0.75 \\[1pt]

% Bloc 60%
\multirow{3}{*}{\cellcolor{yellow!30}\textbf{60\%}}
 & \cellcolor{yellow!30}0.30 & \cellcolor{yellow!30}0.98 & \cellcolor{yellow!30}0.71 & \cellcolor{yellow!30}0.82 & \cellcolor{yellow!30}0.77 \\
 & \cellcolor{yellow!30}0.35 & \cellcolor{yellow!30}0.95 & \cellcolor{yellow!30}0.77 & \cellcolor{yellow!30}0.85 & \cellcolor{yellow!30}0.79 \\
 & \cellcolor{yellow!30}0.40 & \cellcolor{yellow!30}0.92 & \cellcolor{yellow!30}0.82 & \cellcolor{yellow!30}0.87 & \cellcolor{yellow!30}0.81 \\[1pt]

% Bloc 80%
\multirow{3}{*}{\cellcolor{green!30}\textbf{80\%}}
 & \cellcolor{green!30}0.30 & \cellcolor{green!30}0.95 & \cellcolor{green!30}0.70 & \cellcolor{green!30}0.81 & \cellcolor{green!30}0.73 \\
 & \cellcolor{green!30}0.35 & \cellcolor{green!30}0.91 & \cellcolor{green!30}0.75 & \cellcolor{green!30}0.82 & \cellcolor{green!30}0.73 \\
 & \cellcolor{green!30}0.40 & \cellcolor{green!30}0.89 & \cellcolor{green!30}0.78 & \cellcolor{green!30}0.83 & \cellcolor{green!30}0.74 \\
\hline
\end{tabular}
\end{table}

\subsection{Generalization Analysis}
The relevance of a QML/ML model extends beyond its performance on a single dataset; it must also demonstrate the ability to generalize and maintain stable, reliable outcomes when applied in different contexts. In the case of credit card fraud detection, this property is particularly important, as the characteristics of fraudulent activities vary significantly across financial systems, geographical regions, and user behaviors. Consequently, effective models must be validated on multiple datasets to assess their adaptability.  

To evaluate this capacity, we implement the FiD-QAE model on additional credit card fraud datasets representing diverse unbalanced binary classification scenarios. This experimental setup allows us to assess the robustness, stability, and adaptability of FiD-QAE in the presence of structural and statistical variations across datasets. The results are summarized in Table~\ref{comparison_data}.

\begin{table}[htpb]
\centering
\caption{FiD-QAE performance metrics across credit card fraud datasets, showing consistent generalization.}
\begin{tabularx}{\columnwidth}{YYY}
\hline
\rowcolor{cyan!20}
\textbf{Metric} & \textbf{Dataset 2\cite{EalaxiKaggle2017}} & \textbf{Dataset 3\cite{KaggleCreditFraud}} \\
\hline
Number of qubits   & 3   & 4  \\
Trash qubits       & 1   & 1 \\
Accuracy           & 0.82 & 0.83  \\
Precision          & 0.62 & 0.92 \\
Recall             & 0.95 & 0.85 \\
Specificity        & 0.76 & 0.76 \\
F1-score           & 0.75 & 0.88  \\
G-Mean             & 0.85 & 0.8 \\
MCC                & 0.65 & 0.56 \\
\hline
\end{tabularx}
\label{comparison_data}
\end{table}

\subsection{Noise Robustness Analysis}
After establishing a noise-free reference evaluation of the optimized model, we introduced different types of quantum noise to analyze the robustness of FiD-QAE under more realistic conditions. The evaluation considered several common noise channels, including amplitude damping, bit flip, depolarizing, phase damping, and phase flip. In the first stage, each noise type is applied with a probability parameter $p$ varying from 0 to 1 to investigate performance degradation as a function of noise intensity. In the second stage, to isolate the effect of the number of shots on statistical accuracy, the noise probability is fixed at $p=0.5$ for all channels, and the FiD-QAE is evaluated with different shot counts. This two-step process enables direct comparison between noisy and noise-free scenarios and provides insights into both noise resilience and statistical stability.  

\begin{figure}[htpb]
    \centering
    \includegraphics[width=1\linewidth]{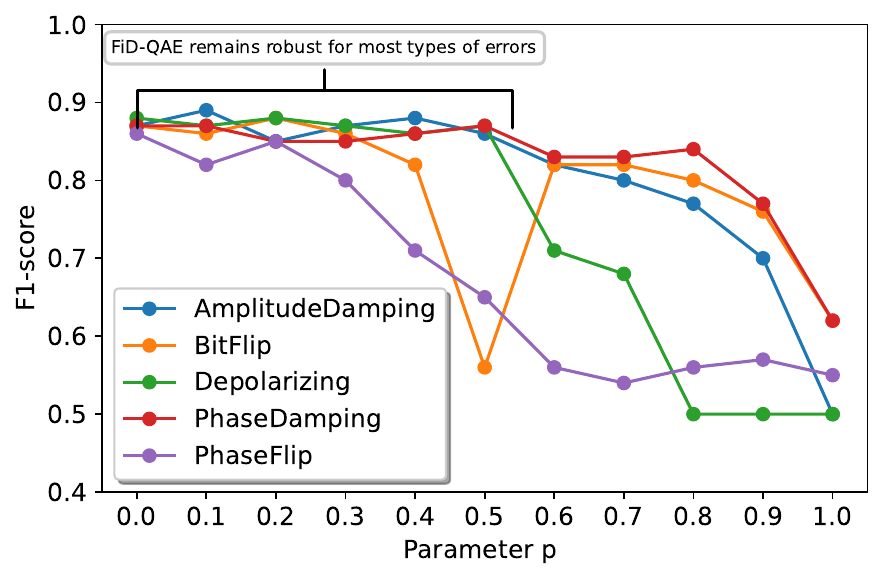}
\caption{Impact of different quantum noise models on the F1-score of FiD-QAE while varying the noise probability $p$.}
    \label{plot_noise}
\end{figure}

\begin{figure}[htpb]
    \centering    
    \includegraphics[width=1\linewidth]{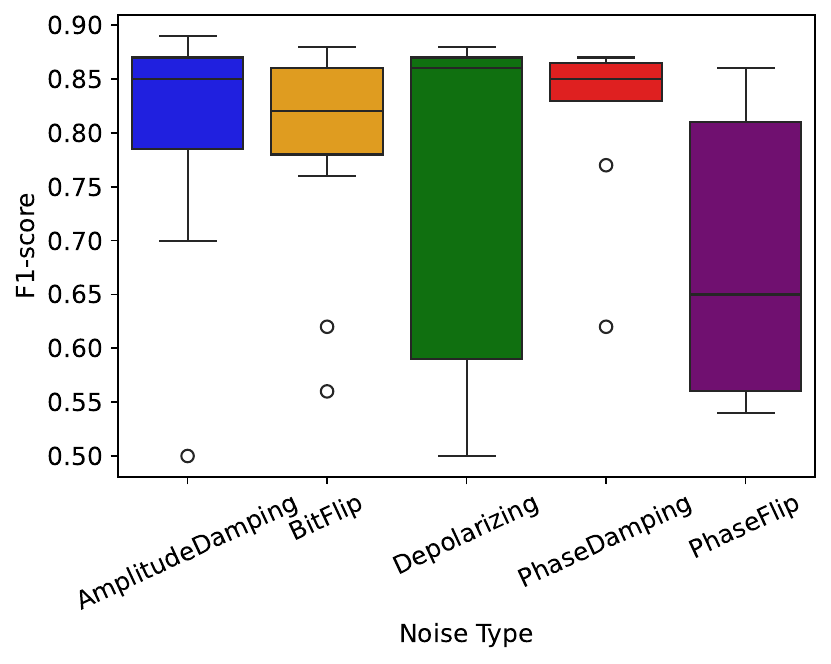}
\caption{Box plots of F1-scores for the FiD-QAE under five quantum noise models. Each box represents the distribution of F1-scores across eleven values of the noise parameter $p$ for the corresponding model.}
    \label{box_noise}
\end{figure}

As shown in Fig.~\ref{plot_noise}, the FiD-QAE demonstrates notable robustness against several noise types. For dissipative channels such as amplitude damping and phase damping, performance remains consistently high across a broad range of noise probabilities, with the F1-score showing significant degradation only when $p > 0.8$. This indicates that FiD-QAE retains reliable predictive power even under conditions of energy loss or partial decoherence. In contrast, the bit flip channel exhibits irregular behavior, with a sharp decline around $p=0.5$ followed by partial recovery, highlighting the uneven effect of this error type. The phase flip channel shows a pronounced deterioration at moderate values of $p$, suggesting that the FiD-QAE is somewhat sensitive to phase reversals, though F1-scores remain at acceptable levels. Finally, the depolarizing channel maintains stability up to $p=0.5$, after which performance declines more irregularly, making it the noise type with the strongest negative impact at higher intensities.  
These observations are further confirmed by Fig.~\ref{box_noise}, which illustrates the distribution of F1-score values for each noise type, complementing the previous analysis by highlighting the variability and stability of FiD-QAE performance. Most distributions are concentrated around high values, with medians between 0.83 and 0.88, indicating strong stability and tolerance to noise. Amplitude Damping, Phase Damping, and Bit Flip noise channels exhibit narrow distributions with medians exceeding 0.85, confirming that FiD-QAE achieves consistent and robust performance under these noise types. In contrast, depolarizing and phase flip noise show greater variability. Nevertheless, depolarizing noise maintains a relatively high median, reflecting resilience despite its severity, while phase flip reveals stronger sensitivity with a lower median of approximately 0.65. These results highlight the robustness of the FiD-QAE model, which continues to achieve competitive F1-scores even under challenging noise conditions.
  
\begin{figure}[htpb]
    \centering
    \includegraphics[width=1\linewidth]{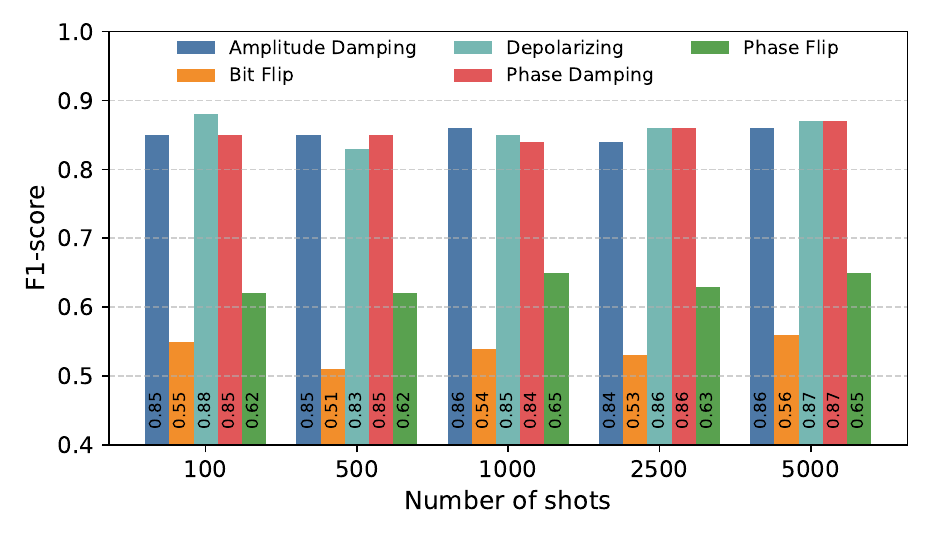}
\caption{Effect of the number of shots and noise type on the F1-score of the FiD-QAE at $p=0.5$.}
    \label{shots}
\end{figure}

To complete the robustness analysis, we also evaluate the effect of the number of shots, which corresponds to the number of measurement repetitions used to estimate output probabilities. This parameter is critical in experimental practice, as it directly influences both statistical accuracy and computational cost on real quantum processors. As shown in Fig.~\ref{shots}, where the noise parameter is fixed at $p=0.5$, the FiD-QAE exhibits remarkable stability with respect to the number of shots. The F1-score remains consistent after only a few hundred repetitions, demonstrating that increasing shots does not yield significant performance gains. This finding indicates that the FiD-QAE efficiently exploits statistical information from quantum measurements and that reliable results can be obtained without resorting to excessively large shot counts, thereby reducing the experimental cost of quantum evaluations.  
These results emphasize the robustness and practicality of FiD-QAE under realistic conditions. The model tolerates a variety of quantum noise sources and maintains stable performance across different shot configurations, underscoring its potential for deployment on current noisy intermediate-scale quantum (NISQ) devices.  
\begin{table*}[htbp]
\centering
\caption{Comparison between FiD-QAE and existing models on the same dataset in terms of performance and reported metrics.}
\label{tab:comparaison_models}
%\begin{tabular}{lccccccccc}              
\begin{tabularx}{\linewidth}{p{2cm}YYYYYYYYY}

\hline
\rowcolor{cyan!20}
\textbf{Model} & \textbf{Qubit} & \textbf{Accuracy} & \textbf{Precision} & \textbf{Recall} & \textbf{Specificity} & \textbf{F1-score} & \textbf{G-Mean} & \textbf{MCC} & \textbf{Metrics} \\
\hline
Classical AE \cite{al2019credit}        & -- & 0.80 & 0.09 & 0.91 & -- & -- & -- & --  & 3 \\
QO-SVM \cite{kyriienko2022unsupervised} & 20  & --  & 0.70 & --  & -- & -- & -- & --  & 1 \\
QGNN \cite{innan2024financial1}         & 6   & 0.92 & 0.94 & 0.79 & -- & 0.86 & -- & --  & 4 \\
QAE-FD \cite{huot2024quantum}           & 4   & 0.99 & 0.37 & 0.89 & -- & 0.53 & -- & --  & 4 \\
\textbf{FiD-QAE}                        & \textbf{4} & \textbf{0.92} & \textbf{0.90} & \textbf{0.83} & \textbf{0.96} & \textbf{0.87} & \textbf{0.89} & \textbf{0.81}  & \textbf{7} \\
\hline
\end{tabularx}
\label{comparison}
\end{table*}
\subsection{Comparison of FiD-QAE with existing models}
We compare the FiD-QAE model with representative approaches that address similar fraud detection problems on the same dataset. While many other studies have explored advanced architectures and alternative evaluation settings, this comparison is restricted to models tested on the same dataset to ensure consistency and fairness.  

As shown in Table~\ref{comparison}, the classical AE achieves high recall (0.91) but very low precision (0.09), resulting in numerous false positives despite an accuracy of 0.80. The QO-SVM reports moderate precision (0.70) but does not provide results for other key metrics and requires 20 qubits, implying higher quantum cost. The QGNN reports accuracy of 0.92, precision of 0.94, and F1-score of 0.86, but recall remains at 0.79. The QAE-FD achieves accuracy of 0.99 and AUC of 0.94, but its imbalance between precision (0.37) and recall (0.89) leads to a relatively low F1-score of 0.53.  

The FiD-QAE achieves accuracy of 0.92, precision of 0.90, recall of 0.83, and F1-score of 0.87. With 4 qubits, FiD-QAE requires the same quantum resources as QAE-FD and fewer than QO-SVM and QGNN. In addition, unlike most prior works, FiD-QAE is evaluated across a broader set of metrics providing a more comprehensive assessment of model behavior under class imbalance.

This comparison should be regarded as a dataset-specific benchmark rather than a comprehensive ranking of all available approaches. The broader range of evaluation metrics reported for FiD-QAE provides additional insight into its performance, complementing prior studies that typically focused on fewer indicators.

\subsection{Hardware Analysis}
% Following a simulation of various types of noise, we conformed our approach on a real quantum processor on order to test the FiD-QAE model against non-idealized hardware defects. To do this, we used the pennylane library coupled with Qiskit Runtime Service interface, which provides access to IBM Quantum's Backend via IBM Cloud.
% We execute our FiD-QAE on the \textit{ibm-torino} processor, with a fixed number of shots set at \hl{...}. The technical configuration relies on calling a remote device with \textit{qiskit.remote} connected to the physical backend using the authentication service provided by IBM Quantum. This choice ensures that the evaluation incorporates real-world constrains such as decoherence, gate error

We conduct a hardware-level evaluation using IBM Quantum Runtime, where both fraud and non-fraud job identifiers are executed and measurement statistics are collected directly from the device (\textit{ibm-torino}). For each job, fidelity (probability of the ideal reference state $s^{*}=\texttt{``000000''}$) and Shannon entropy of the outcome distribution are extracted as discriminative features. These hardware-derived quantities reflect how close the execution is to the target quantum state and how much uncertainty is present in the measurement statistics.

Due to queueing delays, noise accumulation, and financial cost associated with large-scale execution on cloud quantum devices, it is not practical to run exhaustive experiments for every job. Instead, we employ a pragmatic methodology (see Algorithm \ref{alg:hardware_fid}): fidelity–entropy pairs are used as input features to a logistic regression model. The classifier threshold is tuned using Youden’s $J$ statistic on the ROC curve to balance sensitivity and specificity. This hybrid approach leverages the quantum hardware to generate features that encode noise-sensitive quantum information, while relying on a simple classical model to perform the final discrimination. 

The results confirm the feasibility of fidelity-based discrimination under hardware noise. The model achieves an accuracy of $86.6\%$, with a recall of $98.3\%$, ensuring that nearly all fraudulent jobs are flagged. Precision is $79.5\%$, which reflects occasional false alarms due to device fluctuations. The MCC of $0.753$ further confirms strong discriminative capability. While a purely classical logistic regression could also reach high performance, the distinguishing factor here is that the fidelity and entropy features themselves are derived from quantum executions. These hardware-dependent signatures capture aspects of the circuit–device interaction that are not available through classical simulation, making the evaluation an important step toward validating practical quantum workflows. 

\begin{algorithm}[h]
\caption{Fidelity–Entropy Classification on IBM Hardware}
\label{alg:hardware_fid}
\begin{algorithmic}[1]
\Require Job IDs $\{J_i\}$ with labels $y_i$, reference state $s^{*}$
\For{each $J_i$}
    \State Execute $J_i$ on IBM Quantum hardware
    \State Retrieve counts $\{c_k\}$ from measurement outcomes
    \State Compute probabilities $p_k = c_k / \sum_j c_j$
    \State Fidelity: $F_i \gets p_{s^{*}}$
    \State Entropy: $H_i \gets -\sum_k p_k \log_2 p_k$
    \State Store $(F_i, H_i, y_i)$
\EndFor
\State Train logistic regression on $\{(F_i, H_i), y_i\}$
\State Optimize threshold $\tau$ using Youden’s $J$
\State Classify $\hat{y}_i = 1$ if $\hat{p}_i \geq \tau$ else $0$
\State Evaluate metrics (Accuracy, Precision, Recall, F1, MCC)
\end{algorithmic}
\end{algorithm}

\section{Conclusion\label{sec5}}

In this paper, we proposed the FiD-QAE model to enhance credit card fraud detection and address the challenges posed by large and complex datasets. FiD-QAE encodes transactions into quantum states, compresses them into a latent space, and optimizes performance using the SWAP test to assess quantum fidelity, which serves as the central criterion for anomaly detection.  

Extensive experimental evaluation, supported by a wide range of metrics and detailed statistical analyses, demonstrated the robustness of the proposed model. FiD-QAE achieves a balanced trade-off between precision and recall, minimizes false positives, and maintains reliable performance under imbalanced conditions. Sensitivity analyses confirmed the model’s stability across different levels of fraud prevalence and its ability to generalize to new datasets. When compared with existing approaches, FiD-QAE exhibited improved discriminative capability while requiring fewer quantum resources. In addition, the model showed resilience to multiple types of simulated quantum noise, further underlining its suitability for deployment on real quantum hardware, where noise is inevitable.  

This work emphasizes the strategic role quantum models can play in tackling imbalanced classification tasks such as credit card fraud detection. Beyond its methodological contributions, FiD-QAE opens promising directions for advancing quantum autoencoder architectures and exploring their implementation on NISQ devices. These developments have the potential to improve both the reliability and the scalability of financial security systems in real-world settings.  
%\newpage
\section*{Acknowledgment}
This work was supported in part by the NYUAD Center for Quantum and Topological Systems (CQTS), funded by Tamkeen under the NYUAD Research Institute grant CG008.
\bibliographystyle{IEEEtran}

\bibliography{refs}

\end{document}